\documentclass[12pt]{article}

\usepackage{bm}
\usepackage{empheq}
\usepackage{epsfig}
\usepackage{graphicx}
\usepackage{epstopdf}
\usepackage{mathtools}
\usepackage{amssymb}
\usepackage{amsmath}

\makeindex
\def\disp{\displaystyle}

\def\be{\begin{equation}}
\def\ee{\end{equation}}
\def\bea{\begin{eqnarray}}
\def\eea{\end{eqnarray}}
\def\beaN{\begin{eqnarray*}}
\def\eeaN{\end{eqnarray*}}
\def\ed{\end{document}}
\def\bit{\begin{itemize}}
\def\eit{\end{itemize}}

\def\sig{\sigma}
\def\Sig{\Sigma}
\def\lam{\lambda}
\def\Lam{\Lambda}
\def\Del{\Delta}
\def\del{\delta}

\def\k{\kappa}

\def\alf{\alpha}

\def\3nabla{{\stackrel{\scriptscriptstyle 3}{\nabla}}}
\def\2nabla{{\stackrel{\scriptscriptstyle 2}{\nabla}}}

\def\di{\partial}

\def\bn{\bar\nabla}

\def\half{{\textstyle{1 \over 2}}}

\def\~{\tilde}
\def\lag{{{\cal L}}}

\def\m{\label}
\def\l{\left}
\def\r{\right}

\def\goto{\rightarrow}

\def\const{\rm const}

\def\citep{\cite}

\usepackage{cite}

\newcommand{\nc}{\newcommand}

\nc{\bse}{\begin{equation*}}
\nc{\ese}{\end{equation*}}
\nc{\ba}{\begin{array}}
\nc{\ea}{\end{array}}
\nc{\bal}{\begin{align}}
\nc{\eal}{\end{align}}
\nc{\bi}{\begin{description}}
\nc{\ei}{\end{description}}
\nc{\Def}{=}
\nc{\sign}{\mathrm{sign}\;}
\nc{\diagR}{\mathrm{diag}\;}
\nc{\constR}{\mathrm{const}}
\nc{\Tr}{\mathrm{Tr}\,}
\nc{\id}{\mathrm{id}}
\nc{\eq}{\equiv}
\nc{\we}{\wedge}
\nc{\ra}{\rightarrow}
\nc{\bfrac}{\disp\frac}

\nc{\bfpa}{{\bm{\partial}}}                       
\nc{\pa}[1]{{\partial_{#1}}{}}                    
\nc{\pau}[1]{{\partial^{#1}}{}}                   
\nc{\alp}{\alpha}
\nc{\bet}{\beta}
\nc{\gam}{\gamma}
\nc{\eps}{\epsilon}
\nc{\veps}{\varepsilon}
\nc{\zet}{\zeta}
\nc{\tet}{\theta}
\nc{\vtet}{\vartheta}
\nc{\iot}{\iota}
\nc{\kap}{\kappa}
\nc{\vkap}{\varkappa}
\nc{\vpi}{\varpi}
\nc{\vrho}{\varrho}
\nc{\vsig}{\varsigma}
\nc{\ups}{\upsilon}
\nc{\vphi}{\varphi}
\nc{\ome}{\omega}

\nc{\Gam}{\Gamma}
\nc{\Tet}{\Theta}
\nc{\Ups}{\Upsilon}
\nc{\Ome}{\Omega}

\nc{\BL}[1]{\bm{#1}}

\nc{\bfa}{\BL{a}}
\nc{\bfb}{\BL{b}}
\nc{\bfc}{\BL{c}}
\nc{\bfd}{\BL{d}}
\nc{\bfe}{\BL{e}}
\nc{\bff}{\BL{f}}
\nc{\bfg}{\BL{g}}
\nc{\bfh}{\BL{h}}
\nc{\bfi}{\BL{i}}
\nc{\bfj}{\BL{j}}
\nc{\bfk}{\BL{k}}
\nc{\bfl}{\BL{l}}
\nc{\bfm}{\BL{m}}
\nc{\bfn}{\BL{n}}
\nc{\bfo}{\BL{o}}
\nc{\bfp}{\BL{p}}
\nc{\bfq}{\BL{q}}
\nc{\bfr}{\BL{r}}
\nc{\bfs}{\BL{s}}
\nc{\bft}{\BL{t}}
\nc{\bfu}{\BL{u}}
\nc{\bfv}{\BL{v}}
\nc{\bfw}{\BL{w}}
\nc{\bfx}{\BL{x}}
\nc{\bfy}{\BL{y}}
\nc{\bfz}{\BL{z}}

\nc{\bfA}{\BL{A}}
\nc{\bfB}{\BL{B}}
\nc{\bfC}{\BL{C}}
\nc{\bfD}{\BL{D}}
\nc{\bfE}{\BL{E}}
\nc{\bfF}{\BL{F}}
\nc{\bfG}{\BL{G}}
\nc{\bfH}{\BL{H}}
\nc{\bfI}{\BL{I}}
\nc{\bfJ}{\BL{J}}
\nc{\bfK}{\BL{K}}
\nc{\bfL}{\BL{L}}
\nc{\bfM}{\BL{M}}
\nc{\bfN}{\BL{N}}
\nc{\bfO}{\BL{O}}
\nc{\bfP}{\BL{P}}
\nc{\bfQ}{\BL{Q}}
\nc{\bfR}{\BL{R}}
\nc{\bfS}{\BL{S}}
\nc{\bfT}{\BL{T}}
\nc{\bfU}{\BL{U}}
\nc{\bfV}{\BL{V}}
\nc{\bfW}{\BL{W}}
\nc{\bfX}{\BL{X}}
\nc{\bfY}{\BL{Y}}
\nc{\bfZ}{\BL{Z}}

\nc{\bfalp}{\bm{\alp}}
\nc{\bfbet}{\bm{\bet}}
\nc{\bfgam}{\bm{\gam}}
\nc{\bfdel}{\bm{\del}}
\nc{\bfeps}{\bm{\eps}}
\nc{\bfveps}{\bm{\veps}}
\nc{\bfzet}{{\bm{\zet}}}
\nc{\bfeta}{\bm{\eta}}
\nc{\bftet}{\bm{\tet}}
\nc{\bfvtet}{\bm{\vtet}}
\nc{\bfiot}{\bm{\iot}}
\nc{\bfkap}{\bm{\kap}}
\nc{\bfvkap}{\bm{\vkap}}
\nc{\bflam}{\bm{\lam}}
\nc{\bfmu}{\bm{\mu}}
\nc{\bfnu}{\bm{\nu}}
\nc{\bfxi}{\bm{\xi}}
\nc{\bfpi}{\bm{\pi}}
\nc{\bfvpi}{\bm{\vpi}}
\nc{\bfrho}{\bm{\rho}}
\nc{\bfvrho}{\bm{\vrho}}
\nc{\bfsig}{\bm{\sig}}
\nc{\bfvsig}{\bm{\vsig}}
\nc{\bftau}{\bm{\tau}}
\nc{\bfups}{\bm{\ups}}
\nc{\bfphi}{\bm{\phi}}
\nc{\bfvphi}{\bm{\vphi}}
\nc{\bfchi}{\bm{\chi}}
\nc{\bfpsi}{\bm{\psi}}
\nc{\bfome}{\bm{\ome}}

\nc{\bfGam}{\bm{\Gam}}
\nc{\bfDel}{\bm{\Del}}
\nc{\bfTet}{\bm{\Tet}}
\nc{\bfLam}{\bm{\Lam}}
\nc{\bfXi}{\bm{\Xi}}
\nc{\bfPi}{\bm{\Pi}}
\nc{\bfSig}{\bm{\Sig}}
\nc{\bfUps}{\bm{\Ups}}
\nc{\bfPhi}{\bm{\Phi}}
\nc{\bfPsi}{\bm{\Psi}}
\nc{\bfOme}{\bm{\Ome}}

\begin{document}

\centerline{\bf Conserved quantities for black hole solutions }
\centerline{\bf in pure Lovelock gravity}

\smallskip

\centerline{\it A.N.Petrov}

\smallskip

\centerline{\it Moscow MV Lomonosov State university, Sternberg
Astronomical institute,} \centerline{\it
 Universitetskii pr., 13, Moscow, 119992,
RUSSIA }

\smallskip

\centerline{ e-mail: alex.petrov55@gmail.com}

\smallskip

telephone number: +7(495)7315222

\smallskip

fax: +7(495)9328841

\smallskip


\begin{abstract}
We construct conserved quantities in pure Lovelock gravity for both static and dynamic Vaydia-type black holes with AdS, dS and flat asymptotics, applying field-theoretical formalism developed earlier. Global energy (where applicable), quasi-local energy together with fluxes of these quantities are presented for both types of black holes, considering asymptotic spacetime as background. The same quantities are constructed for dynamic black holes on the background of the related static black holes. Besides, for the dynamic black holes, energy densities and densities of energy flux are calculated in the frame of freely and radially falling observer on the background of the related static black holes. All the constructed energetic characteristics are analyzed and discussed in detail.
\end{abstract}

\section{Introduction}
\m{Introduction}

The Lovelock theory \cite{Lovelock} is possibly the most popular one amongst multidimensional gravitational theories. The Lovelock gravity has the following advantages 1) it is a {\em covariant} metric theory in $D$ dimensional spacetime; 2) its field equations are of the {\em second} order only; 3) Einstein-Gauss-Bonnet theory, being an example of second order of Lovelock gravity, presents the low energy limit of the heterotic string theory \cite{Gross_Witten_1986,Bento_Bertolami_1996}.

The Lovelock Lagrangian is presented by a sum of polynomials constructed from the  Riemannian tensor with each of the polynomials multiplied with a corresponding coupling constant. Due to the quite complicated mathematical structure it is impossible to find a general solution except for simple cases. Additional obstacle in the most general case is that it is not possible to give physical interpretation for each of the constants. Therefore, it is necessary to reduce the number of the coupling constants. One of the popular ways is to consider the so-called {\em pure} Lovelock gravity where one has only a zeroth term (a bare cosmological constant) and a single Lovelock polynomial with a related coupling constant.

Black hole solutions in the pure Lovelock gravity have initially been suggested in the work \cite{Troncoso+_2000}. Such black holes have flat asymptotics and are related to Lagrangians where the zeroth term is equal to zero. In the works \cite{Kastor_Mann_2006} and \cite{Giribet_Oliva_Troncoso_2006}, these black hole solutions were extended to asymptotically flat black string solutions and black brane solutions,  respectively. New wider families of spherically symmetric solutions have been suggested in the works by Cai et. al. (see \cite{Cai_Ohta_2006} for static solutions and \cite{Cai+_2008} for generalized Vaidya-like solutions).

Properties of the pure Lovelock gravity are studied in the series of works by Dadhich and coauthors. In \cite{Dadhich_2010}, a special structure of the Lagrangian, field equations and conserved expressions are considered and related formalism of their applications is developed.
In \cite{Dadhich+_2012a,Dadhich+_2013a,Dadhich+_2013b,Dadhich_2016} the formalism of \cite{Dadhich_2010} is used to construct and examine solutions of the pure Lovelock gravity in critical odd and even dimensions, which describe the field of a global monopole, collapse and boundary orbits.  In \cite{Dadhich_2011,Dadhich+_2012,Dadhich+_2013} spherically symmetric static black holes and their thermodynamics are studied. The interest in the pure Lovelock gravity continues to be very high, see, for example, \cite{F_2020,Dadhich+_2020,Dadhich+_2020a,Kastikainen_2020}.

After noting the attractive features of the pure Lovelock gravity mentioned above, it is also necessary to note its problems. For example, the presence of cubic kinetic terms and quadratic constraints in 5-dimensional Lanczos-Lovelock theory makes it intrinsically nonlinear \cite{Deser_Franklin}. Indeed, Lagrangian of a linearized version of the theory, which should be quadratic in perturbations, becomes a divergence. Thus, the ``linearized'' version has a cubic Lagrangian rather than quadratic one. The attempts to remove this pathology have not been particularly successful.

To the best of our knowledge, the efforts of constructing and studying conserved quantities in the pure Lovelock gravity, especially for dynamic black holes, have not been exhaustive. One of the reasons is that the researchers did not use the more advanced mathematical technique which is currently available. This technique has been worked out in our work \cite{Petrov_2019}, where we have constructed conserved quantities for perturbations in the Lovelock gravity in the most general form. Constructions in \cite{Petrov_2019} have been obtained on the basis of the field-theoretical approach (see \cite{GPP,[15],GP87,CQG_DT,Petrov_2008,Petrov+_2017}, the recent review \cite{Petrov_Pitts_2019}, and references therein). The formalism has the following important properties: 1) it presents conserved currents and related superpotentials for perturbations on {\em arbitrary} curved  backgrounds not only on maximally symmetric ones, like anti-de Sitter (AdS), de Sitter (dS) or flat backgrounds; 2) it is covariant and exact, not approximate; 3) it allows us to make the use of {\em arbitrary} displacement vectors, not restricted to Killing vectors of a background (these could be proper vectors of various observers, conformal Killing vectors, etc). In this relation we would remark the recent work \cite{Krishna_2019} where the conserved Abbott-Deser-Tekin \cite{DT1,DT2,DT_2019} currents and potentials are generalized to the presentation with  non-Killing displacement vectors in the Lovelock gravity.

The main goal of the present paper is to apply formalism \cite{Petrov_2019} (adopted to pure Lovelock gravity) to {\em construct conserved quantities} for both, the static \cite{Cai_Ohta_2006} and the dynamic Vaydia-like \cite{Cai+_2008} black holes with AdS, dS and flat asymptotics. The listed above properties of the field-theoretical method allows us to construct
\bit

\item[1)] global and quasi-local energy for both dynamic and static black holes on the AdS, dS and flat backgrounds;

\item[2)] global and quasi-local energy  for dynamic black holes on the backgrounds of related static solutions;

\item[3)] for dynamic solutions, the energy densities and the energy flux densities measured by freely and radially falling observers on a background of related static geometries.

\eit
Thereby, we plan to close the aforementioned gap in constructing conserved quantities, at least partially.

It is worth mentioning that we will study dynamic and related static black holes as unique complexes. Indeed, the same physical object can be considered both in static and dynamic regimes in different periods of time.\footnote{This idea is not a new one, for example, in \cite{Nozawa_Maeda_2006}, in the framework of dimensionally continued Lovelock gravity the authors analyzed formation of Banados-Teitelboim-Zanelli black holes and naked singularities during restricted periods of time when null fluid collapses.} Such an approach allows us to study dynamic solutions as perturbations of the corresponding static solutions. In this case the conserved quantities can be defined on a basis of observers connected with the frame of static solutions, so that their interpretation becomes more meaningful.

The paper is organized as follows. In section \ref{Preliminaries}, we 1) focus on the structure, for example, of the generalized energy-momentum of the light-like fluid and discuss necessary properties of the solutions \cite{Cai_Ohta_2006} and \cite{Cai+_2008}; 2) introduce the basic concepts of the field-theoretical approach and explain the method of construction of conserved currents and superpotentials in an arbitrary metric theory; 3) adopt superpotential expressions constructed in \cite{Petrov_2019} to the case of pure Lovelock gravity.

In section \ref{C_Q_Lovelock}, 1) the adopted formulae of \cite{Petrov_2019} are represented for black hole metrics of the Vaidya type in ingoing coordinates; 2) all the non-zero components of conserved currents and superpotentials, expressions for charges and proper vectors of freely and radially falling observers in the general form are presented.

In sections \ref{BH_AdS}, \ref{BH_dS} and \ref{BH_flat}, we study static and Vaidya-like black holes with AdS, dS and flat asymptotics, respectively. We calculate global and quasi-local energy and related fluxes in correspondence with  the main goal of the current paper that were formulated above in points 1) - 3).

In section \ref{Discussion}, we systemize and discuss our findings.

In Appendix, we present lengthy expressions for the energy densities and densities of energy flux measured by freely falling observers.

\section{Preliminaries}
\setcounter{equation}{0}
\m{Preliminaries}

\subsection{The pure Lovelock gravity and Vaidya-like solutions. }
 \m{Lovelock}
\setcounter{equation}{0}

Lagrangian for the Lovelock theory \cite{Lovelock} adopted to the pure Lovelock gravity is given by
 \bea
 \lag(g,\Phi) &=& - \frac{1}{2\k}\lag_{\ell}(g_{\mu\nu}) + \lag_{m}(g_{\mu\nu},\Phi^A)\,,
 \m{lag-ll}\\
 \lag_{\ell}&\equiv&\sqrt{-g}
\l( \alf_0 + \frac{\alf_p}{2^p} \delta _{\lbrack \mu_{1}\mu_{2}\cdots \mu_{2p}]}^{[\nu_{1}\nu_{2}\cdots \nu_{2p}]}\,{R}_{\nu_{1}\nu_{2}}^{\mu_{1}i_{2}}\cdots {R}_{\nu_{2p-1}\nu_{2p}}^{\mu_{2p-1}\mu_{2p}}\r)  \,,
\label{Lovelock-lag+}
 \eea
where only the coupling constants $\alf_0$ and $\alf_p$ are included. Here, $\k = 2\Omega_{D-2}G_D> 0$  with
$D$-dimensional Newton's constant $G_D$; $\Omega_{D-2}$ is a square of $(D-2)$-dimensional sphere with the unit radius; $\Phi^A$ is a generalized notation for matter fields; $\delta _{\left[ \mu _{1}\cdots \mu _{q}\right] }^{\left[ \nu _{1}\cdots \nu_{q}\right] }$ is the totally-antisymmetric Kronecker delta.
The field equations related to (\ref{lag-ll})  and (\ref{Lovelock-lag+}) are
\begin{equation}
\frac{\delta\lag_{\ell}}{\delta g_{\rho\sig}}= \sqrt{-g}\l(g^{\rho\sig}\frac{\alf_0}{2} +g^{\pi\rho}
\frac{\alf_p}{2^{p+1}} \delta _{\lbrack \pi \nu_{1}\nu_{2}\cdots
\nu_{2p}]}^{[\sig \mu_{1}\mu_{2}\cdots \mu_{2p}]}\,{R}_{\mu_{1}\mu_{2}}^{\nu_{1}\nu_{2}}\cdots {R}_{\mu_{2p-1}\mu_{2p}}^{\nu_{2p-1}\nu_{2p}}\r)  = - \kappa \sqrt{-g}{T}^{\rho\sig}   \label{Lovelock-eqs+}
\end{equation}
with the matter energy-momentum $T_{\alf\beta} =
{2}({\sqrt{-g}})^{-1}{\delta\lag_m}/{\delta g^{\alf\beta} }$. For $p > \l[(D-1)/2\r]$ Lagrangian $\lag_{\ell}$ does not contribute to the Euler-Lagrange equations (\ref{Lovelock-eqs+}); `bare' cosmological constant can be either $\alf_0 \neq 0$ or $\alf_0 = 0$.

Following  \cite{Cai+_2008}, we study  solutions to the equations (\ref{Lovelock-eqs+}) presented by the metric of the Vaidya form in ingoing coordinates:
\be
ds^2 = -f(r,v)dv^2 +2dvdr + r^2q_{ij}dx^idx^j
\,
\m{metric}
\ee
with the generalized energy-momentum for the null fluid. The advanced null coordinate $v$ is numerated as $v=x^0$; the radial coordinate $r=x^1$; $q_{ij}$ is a metric on the unit $(D-2)$-dimensional sphere. To derive solutions to the equations (\ref{Lovelock-eqs+}) in the succinct form it is convenient to re-denote coupling constants, as in \cite{Cai_Ohta_2006}:
\be
\frac{\alf_0}{(D-1)(D-2)} \equiv -\frac{1}{\ell^2}, \qquad
\frac{\alf_p (D-3!)}{(D-2p-1)!} \equiv \alf^{2p-2}
\,.
\m{coupling}
\ee

From the start to construct the energy-momentum for the null fluid one assumes, as usual \cite{Cai+_2008}, that only the component $T_0{}^1$ in (\ref{Lovelock-eqs+}) in the case of the metric (\ref{metric}) is not equal to zero. Then after integrating a combination of the equation (\ref{Lovelock-eqs+}) components $\{_0{}^0\}$ and $\{_i{}^k\}$ one includes into consideration, $m(v)$, an arbitrary function of $v$.
To introduce the {\em generalized} energy-momentum for null fluid the authors \cite{Cai+_2008} (see also \cite{Dominguez_Gallo_2006}) assume $T_0{}^0 =T_1{}^1 \neq 0$. Then the consistency condition $\nabla_\beta T_{\alf}{}^{\beta} = 0$ is integrated to obtain  $T_0{}^0 =T_1{}^1 = {\cal C}(v)r^{-(D-1)(1-\sigma)}$ where $\nabla_\beta$ is a covariant derivative with respect to $g_{\mu\nu}$ and ${\cal C}(v)$ is another arbitrary function of $v$. The natural interpretation of non-zero $T_1{}^1$ is a radial pressure $P_r = - T_1{}^1 = -{\cal C}(v)r^{-(D-1)(1-\sigma)}$, thus, non-zero ${\cal C}(v)$ signals the presence of $P_r$ and defines its time evolution. In addition, a tangential pressure $P_\theta = -\sigma P_r$ is assumed with an arbitrary constant $\sigma$. Finally, the generalized energy-momentum becomes
\be
{T}_{\alf\beta} = \mu\l( n_\alf n_\beta \r)+ P_r \l(n_\alf l_\beta + l_\alf n_\beta \r) + P_\theta r^2q_{\alf\beta}\,,
\m{T_light}
\ee
where ingoing and outgoing null vectors are $n^\alf = \l(0,-1, 0,\ldots, 0\r)$ and $l^\alf = \l(1,f/2, 0,\ldots, 0\r)$, respectively,
for which $n^\alf n_\alf = 0$, $l^\alf l_\alf = 0$, $l^\alf n_\alf = -1$. The energy density
\be
 \mu = \frac{1}{\k}\l( \frac{\dot m(v)}{r^{D-2}} + \frac{\dot {\cal C}(v)\Theta (r) }{r^{D-2}}\r)\,
\m{mu_light}
\ee
is obtained by the requirement of the consistency of the system (\ref{Lovelock-eqs+}) and defines the source of its  $\{_0{}^1\}$-component. $\Theta (r)$ is a result of integration when $\mu$ is constructed.
The dominant energy conditions applied in \cite{Dominguez_Gallo_2006} to the energy-momentum (\ref{T_light}) presenting the null fluid of the type $II$ in classification of \cite{Hawking_Ellis_1973} give
$\mu \geq 0$,  ${\cal C}(v) \le 0$ and $-1 \le \sig \le 0$. More details for $\Theta (r)$ and $\sig$ are
\bea
&&\sig = -1/(D-2) ~~\goto ~~ \Theta(r) = \ln(r)\,;
\m{Theta_1}\\
&&-1 \le \sig \le 0 ~~ {\rm with}~~\sig \neq -1/(D-2)  ~~\goto ~~  \Theta(r) = \frac{r^{(D-2)\sig +1}}{(D-2)\sig +1}\,.
\m{Theta_2}
\eea

In \cite{Cai+_2008}, the equations (\ref{Lovelock-eqs+}) are resolved for the even and odd $p$, respectively, as
\bea
f(r,v) &=& 1 \pm \frac{r^2}{\alf^{2-2/p}}\l[\frac{1}{\ell^2} + \frac{ m_0 + m(v)}{r^{D-1}} + \frac{{\cal C}(v)\Theta (r) }{r^{D-1}}\r]^{1/p}
\,,\m{even+}\\
f(r,v) &=& 1 - \frac{r^2}{\alf^{2-2/p}}\l(\frac{1}{\ell^2} + \frac{ m_0 + m(v)}{r^{D-1}} + \frac{{\cal C}(v)\Theta (r) }{r^{D-1}}\r)^{1/p}
\,.
\m{odd+}
\eea
In the case $m(v) = {\cal C}(v) = 0$ the solutions (\ref{even+}) and (\ref{odd+}) are simplified to the related static solutions in \cite{Cai_Ohta_2006}\footnote{The static solutions  in \cite{Cai_Ohta_2006} have been derived in the Schwarzschild-like coordinates: $t,r,\theta,\phi, \ldots$. Here, for the sake of uniqueness we use the form (\ref{metric}) for the static solutions with $f(r,v) = f(r)$ in the Eddington-Finkelstein coordinates. To obtain the form (\ref{metric}) with $f(r)$ from the solutions given in \cite{Cai_Ohta_2006} the coordinate transformation $dt = dv + dr/f(r)$ is applied. } with the constant mass parameter $m_0$. The cases $1/\ell^2<0$, $1/\ell^2>0$ and $1/\ell^2=0$ correspond to solutions with AdS, dS and flat asymptotics, respectively.

We only study solutions (\ref{even+}) and (\ref{odd+}), which present black holes only that looks physical \cite{Cai+_2008}, and do not consider other solutions corresponding to naked singularities, etc. To be specific, we restrict ourselves by the following natural requirements: 1) the constant mass parameter, defining static black holes, satisfies $m_0>0$; 2) a presence of radiating matter is defined by $\mu>0$, it falls in a restricted interval $0\le v \le v_0$; 3) then, from (\ref{mu_light}), it follows that the sum $ m(v) + {\cal C}(v)\Theta (r)$ is increasing with $v$ and therefore corresponds to {\em monotonic evolution};\footnote{We do not consider $m(v)$ and ${\cal C}(v)$ separately, as this is sufficient for purposes of the present paper. Some variants of relations between $m(v)$ and ${\cal C}(v)$ in the Einstein-Gauss-Bonnet gravity can be found in \cite{Dominguez_Gallo_2006}.}
 4) lastly, we consider $D \ge 2p+2$ only.

\subsection{The field-theoretical approach in metric theories}
\m{ON_FT_approach}

Let us present the main notions of the field-theoretical method in constructing conserved quantities in an arbitrary metric theory. Consider Lagrangian for such a theory:
\be
 \lag(g_{\mu\nu},\Phi^A) = - \frac{1}{2\k}\lag_{g}(g_{\mu\nu}) + \lag_{m}(g_{\mu\nu},\Phi^A)\,
 \m{lag-g}
 \ee
which depends on the first and second  derivatives of $g_{\mu\nu}$ only. The related Euler-Lagrange equations are of the second order, in analogy to the Lovelock theory.
A physical system presented by the Lagrangian (\ref{lag-g}) can be considered as a perturbed one
with respect to a background system with the Lagrangian $\bar\lag=\lag(\bar g_{\mu\nu},\bar\Phi^A)$ where `bar' means a background quantity. In the present paper we consider vacuum backgrounds only, therefore  $\bar\lag=\lag_g(\bar g_{\mu\nu})$ and $\bar\Phi^A=0$. We
decompose metric variables in (\ref{lag-g}) into the
background parts and the dynamic variables (perturbations):
\be
 g_{\alf\beta} = \bar g_{\alf\beta} + \varkappa_{\alf\beta} \,,
\m{varkappa}
\ee
considering the matter variables as dynamic variables (perturbations) $\Phi^A = \bar\Phi^A+\phi^A= \phi^A$ in the background spacetime with the metric $\bar g_{\mu\nu}$.

The perturbed system can be presented by the Lagrangian \cite{GPP,[15]}:
 \be
\lag^{dyn}(\bar g;\,\varkappa,\phi) = \lag (\bar g+\varkappa,\,\phi ) - \varkappa_{\mu\nu} \frac{\delta \bar \lag_g}{\delta \bar g_{\mu\nu}} - \bar \lag_g\,.
 \m{lag}
 \ee
We call it the {\em dynamic Lagrangian} because perturbations $\varkappa_{\mu\nu}$ and $\phi^A$ in (\ref{lag}) are treated as {\em dynamic variables}. It is important to note that for small $\varkappa_{\mu\nu}$ expression (\ref{lag}) in general rather reduces to the Lagrangian of second order in perturbations that presents a linearized version of the theory.

After varying the Lagrangian (\ref{lag}) with respect to $\varkappa_{\mu\nu}$ and algebraic transformations one obtains the gravitational equations in the field-theoretical form:
\be
{\cal G}^{L}_{\mu\nu} = \k{\bm t}_{\mu\nu}\,.
 \m{PERTmunu}
 \ee
One can show that these equations are equivalent to the field equations of the original theory with the Lagrangian (\ref{lag-g}) if the background equations for $\bar g_{\mu\nu}$ are taken into account. The linear operator on the
left hand side of (\ref{PERTmunu}) and the total symmetric
(metric) energy-momentum tensor density for the fields $\varkappa_{\mu\nu}$ and $\phi^A$ on the
right hand side of (\ref{PERTmunu}) are defined as
 \be
 {\cal G}^{L}_{\mu\nu} =
 \frac{\delta }{\delta \bar g^{\mu\nu}} \varkappa_{\alf\beta}
\frac{\delta \bar \lag_{g}}{\delta \bar { g}_{\alf\beta}}\,,\qquad {\bm t}_{\mu\nu} = 2\frac{\delta\lag^{dyn}}{\delta \bar
 g^{\mu\nu}} = {\bm t}^g_{\mu\nu} + {\bm t}^m_{\mu\nu}\,.
 \m{GL-q}
 \ee
With regards to this expression, it is worth noting that $ \bn_\nu {\cal G}_{L}^{\mu\nu}\equiv 0$ (where $\bn_\nu$ is a covariant derivative with respect to $\bar { g}_{\alf\beta}$); ${\bm t}^g_{\mu\nu}$  is the energy-momentum related to a pure gravitational part of the Lagrangian (\ref{lag}), and ${\bm t}^m_{\mu\nu}$ is the energy-momentum of matter fields $\phi^A$ in (\ref{lag}).

We consider Lagrangian based theories when Lagrangians are scalar densities. Therefore, to derive conservation laws one can use the diffeomorphism invariance and apply the Noether theorem. Following this prescription Lie displacements with related arbitrary enough smooth vectors $\xi^\alf$ are provided. Because initial vectors $\xi^\alf$ are not determined, interpretation of  resulting conserved quantities is unclear. It becomes meaningful when vectors $\xi^\alf$ are chosen in correspondence with a problem under consideration. They could be, for example, Killing vectors, proper vectors of observers, etc.

Here, to construct conserved currents and superpotentials one can apply the Noether procedure to the system (\ref{lag}) directly. However, there is a more concise way \cite{Petrov+_2017,Petrov_Pitts_2019}. Let us consider
\be
\lag_1 = -\frac{1}{2\k}\varkappa_{\alf\beta} \frac{\delta \bar
\lag_{g}}{\delta \bar {g}_{\alf\beta}}\,
 \m{Lag-1}
 \ee
that is a part of $\lag^{dyn}$ defined in (\ref{lag}). Because $\lag_1$ is a scalar density one can apply the Noether theorem as well \cite{Petrov+_2017,Petrov_Pitts_2019}. Doing so and making the use of the  field equations (\ref{PERTmunu}) one obtains the conservation law $\di_\mu{\cal I}^{\mu}(\xi) \equiv \bn_\mu{\cal I}^{\mu}(\xi) = 0$ where the current ${\cal I}^{\mu}(\xi)$ is a vector density. It depends essentially on the energy momentum ${\bm t}_{\mu\nu}$ defined in (\ref{GL-q}). However, for the purposes of the present paper, we are not specifically interested in the structure of ${\cal I}^{\mu}(\xi)$. Instead of this, we use the structure of the superpotential ${\cal I}^{\mu\nu}(\xi)$, antisymmetric tensor density in (\ref{C_Law}), and calculate components of the current as a divergence of ${\cal I}^{\mu\nu}(\xi)$:
\be
{\cal I}^{\mu}(\xi) = \di_\nu{\cal I}^{\mu\nu}(\xi)\equiv \bn_\nu{\cal I}^{\mu\nu}(\xi); \qquad
{\cal I}^{\mu\nu} = {\textstyle{4\over 3}}\l(
 2\xi_\sig \bar\nabla_\lam  {\bm \omega}_{1}^{\sig[\mu|\nu]\lam}  -
{\bm \omega}_{ 1}^{\sig[\mu|\nu]\lam}
\bar \nabla_\lam  \xi_\sig\r)\,.
 \m{C_Law}
\ee
Here, $\xi^\mu$ is an arbitrary displacement vector; the quantity ${\bm \omega}_{1}^{\rho\lam|\mu\nu}$ is defined as
\be
 {\bm \omega}^{\rho\lam|\mu\nu}_1  =   \frac{\di \lag_1}{\di \bar g_{\rho\lam,\mu\nu}}\,; \qquad {\bm \omega}^{\rho\lam|\mu\nu}_1  = {\bm \omega}^{\lam\rho|\mu\nu}_1  =
 {\bm \omega}^{\rho\lam|\nu\mu}_1
 \m{NL1}
 \ee
 with $\lag_1$ given in (\ref{Lag-1}); ${\bm \omega}_{1}^{\rho[\lam|\mu]\nu}$ is antisymmetric in $\lam$ and $\mu$ part of (\ref{NL1}).

\subsection{The field-theoretical superpotentials in pure Lovelock gravity}

In \cite{Petrov_2019} the superpotentials defined in (\ref{C_Law}) have been constructed in the case of the Lovelock gravity in general formulation. Let us reformulate it for the specific case of pure Lovelock gravity. For the background version of Lagrangian (\ref{Lovelock-lag+}), $\bar\lag_{\ell}$, and for the background operator in (\ref{Lovelock-eqs+}), ${\delta\bar\lag_{\ell}}/{\delta \bar g_{\rho\sig}}$,  we define auxiliary Lagrangian (\ref{Lag-1}):
\begin{equation}
\lag_{\ell 1}= - \frac{1}{2\k}\varkappa_{\rho\sig}\frac{\delta\bar\lag_{\ell}}{\delta \bar g_{\rho\sig}}=- \frac{\sqrt{-\bar g}}{2\k} \l(\varkappa^\alf_\alf \frac{\alf_0}{2} + \varkappa_\sig^\rho
\frac{\alf_p}{2^{p+1}} \delta _{\lbrack \rho \nu_{1}\nu_{2}\cdots
\nu_{2p}]}^{[\sig \mu_{1}\mu_{2}\cdots \mu_{2p}]}\,{\bar R}_{\mu_{1}\mu_{2}}^{\nu_{1}\nu_{2}}\cdots {\bar R}_{\mu_{2p-1}\mu_{2p}}^{\nu_{2p-1}\nu_{2p}}\r)   \,.  \label{lock-1}
\end{equation}
Then, the expression (\ref{NL1}) becomes
\bea
&&{\bm \omega}^{\rho\lambda|\mu\nu}_{\ell 1}=    \frac{\di \lag_{\ell 1}}{\di \bar g_{\rho\lam,\mu\nu}}= - \frac{\sqrt{-\bar g}}{2\k} \varkappa^\alf_\beta
\frac{p\alf_p}{2^{p+1}} \delta _{\lbrack \alf\phi\psi \nu_{3}\nu_{4}\cdots
\nu_{2p}]}^{[\beta\pi\sigma \mu_{3}\mu_{4}\cdots \mu_{2p}]}\,{\bar R}_{\mu_{3}\mu_{4}}^{\nu_{3}\nu_{4}}\cdots {\bar R}_{\mu_{2p-1}\mu_{2p}}^{\nu_{2p-1}\nu_{2p}}\bar g^{\phi\tau}\bar g^{\psi\kappa}D^{\rho\lambda\mu\nu}_{\pi\sigma\tau\kappa}  \,; \nonumber \\
&&D^{\rho\lambda\mu\nu}_{\pi\sigma\tau\kappa} \equiv  \half \l(\delta^\rho_\pi\delta^\lambda_\k +\delta^\rho_\k\delta^\lambda_\pi \r)\l(\delta^\mu_\sigma\delta^\nu_\tau +\delta^\mu_\tau\delta^\nu_\sigma \r) \, \label{omega}
\eea
and superpotential in (\ref{C_Law}) in the pure Lovelock gravity acquires the form:
\be
{\cal I}^{\mu\nu} = {\textstyle{4\over 3}}\l(
 2\xi_\sig \bar\nabla_\lam  {\bm \omega}_{\ell 1}^{\sig[\mu|\nu]\lam}  -
{\bm \omega}_{\ell 1}^{\sig[\mu|\nu]\lam}
\bar \nabla_\lam  \xi_\sig\r)\,.
 \m{Super-l}
 \ee

 \section{Conserved quantities for solutions of the Vaidya type}
 \m{C_Q_Lovelock}
\setcounter{equation}{0}

In this section, we adopt the results of \cite{Petrov_2019} to the case of pure Lovelock gravity when the solutions of the type (\ref{metric}) are considered as perturbed ones with respect to static solutions of the same type in the Eddington-Finkelstein coordinates:
\be
d\bar s^2 = -\bar f(r)dv^2 +2dvdr + r^2q_{ij}dx^idx^j
\,.
\m{metric_bar}
\ee
Thus, the non-zero components of metric perturbations defined by (\ref{varkappa}) are
\be
\varkappa_{00} = - f + \bar f = -\Delta f;\qquad \varkappa_{0}^1 = \varkappa^{11} = \varkappa_{00}\,.
\m{varkappa_bar}
\ee
Non-zero background Riemannian tensor components related to (\ref{metric_bar}) are
\be
\bar R^{01}_{01} = -\frac{\bar f''}{2}, \qquad \bar R^{0i}_{0j} = -\frac{\bar f'}{2r}\delta^i_j, \qquad \bar R^{1i}_{1j} = -\frac{\bar f'}{2r}\delta^i_j, \qquad
\bar R^{ij}_{kl} = \frac{1-\bar f}{r^2}\delta^{[ij]}_{[kl]}
\,,
\m{Riemannian_bar}
\ee
where `prime' means $d/dr$.

Let us derive the explicit expression for the quantity ${\bm \omega}^{\rho\lambda|\mu\nu}_{\ell 1}$ defined in (\ref{omega}). In summation (\ref{omega}), only the components $\delta _{\lbrack 1\phi\psi \nu_{3}\nu_{4}\cdots \nu_{2p}]}^{[0\pi\sigma \mu_{3}\mu_{4}\cdots \mu_{2p}]}$ of the totally antisymmetric Kronecker symbol participate because among components $\varkappa_{\alf}^\beta$ in (\ref{varkappa_bar}) the unique component $\varkappa_{0}^1$ is non-zero. This means that only the components $\bar R^{ij}_{kl}$ in (\ref{Riemannian_bar}) contribute to final summation in (\ref{omega}). Therefore, to simplify calculations, without changing the final result, we can replace the components (\ref{Riemannian_bar}) in (\ref{omega}) with $\bar R^{\alf\beta}_{\gamma\delta} = ({1-\bar f})r^{-2}\delta^{[\alf\beta]}_{[\gamma\delta]}$.
Substituting  these components into (\ref{omega}) and proceeding step by step using the relation,
\be
\delta _{\lbrack \beta_{1}\beta_{2}\cdots
\beta_{m}]}^{[\alf_{1}\alf_{2}\cdots \alf_{m}]} \delta^{[\beta_{m-1}\beta_{m}]}_{[\alf_{m-1}\alf_{m}]} = 2[D-(m-1)][D - (m-2)]\delta _{\lbrack \beta_{1}\beta_{2}\cdots
\beta_{m-2}]}^{[\alf_{1}\alf_{2}\cdots \alf_{m-2}]},
\m{+delta+}
\ee
one obtains finally
\be
{\bm \omega}^{\rho\lambda|\mu\nu}_{\ell 1}= - \frac{\sqrt{-\bar g}}{2\k} \varkappa^\alf_\beta
\frac{p\alf_p}{2^{p+1}} \l(\frac{1-\bar f}{r^2} \r)^{p-1} 2^{p-1} \frac{(D-3)!}{(D-2p-1)!}\delta _{\lbrack \alf\phi\psi]}^{[\beta\pi\sigma ]}\,\bar g^{\phi\tau}\bar g^{\psi\kappa}D^{\rho\lambda\mu\nu}_{\pi\sigma\tau\kappa}  \,.  \label{omega++}
\ee
 Recalling the definitions for the coupling constants in (\ref{coupling}) and for $D^{\rho\lambda\mu\nu}_{\pi\sigma\tau\kappa}$ in (\ref{omega}), keeping in mind nonzero components (\ref{varkappa_bar}) and the symmetry properties in (\ref{NL1}), we obtain non-zero components for (\ref{omega++}):
\be
{\bm \omega}^{11|ij}_{\ell 1}= -2 {\bm \omega}^{i1|1j}_{\ell 1} = \frac{\sqrt{-\bar g}}{2\k} \frac{\Delta f}{r^2}\frac{p}{2}\, \l[\frac{\alf^2}{r^2}(1-\bar f) \r]^{p-1}\,.
\m{www}
\ee

The most permissible form for displacement vectors preserving spherical symmetry is
\be
\xi^\alf = \l\{\xi^0(v,r), \xi^1(v,r),0,\ldots,0\r\}; \qquad \xi_\alf = \l\{\xi_0(v,r), \xi_1(v,r),0,\ldots,0\r\} \,.
\label{xi}
\ee
Then, the components for the superpotential (\ref{Super-l}) with (\ref{www}) are as follows:
\bea
{\cal I}^{01}(\xi) &=& - {\cal I}^{10}(\xi) =
 (D-2) \frac{p\sqrt{q}}{2\k} r^{D-3}\Delta f \l[\frac{\alf^2}{r^2}(1-\bar f) \r]^{p-1}\!\!\xi_1(v,r) ,\nonumber\\
{\cal I}^{0i}(\xi) &=&  {\cal I}^{1i}(\xi) = {\cal I}^{ij}(\xi) = 0.
\label{sup_tau}
\eea
This allows us to calculate non-zero components of the current by making use of the conservation law of the general type (\ref{C_Law}):
\bea
{\cal I}^{0}(\xi) &=& \di_1 {\cal I}^{01}(\xi) ,\label{01_CL}\\
{\cal I}^{1}(\xi) &=& \di_0 {\cal I}^{10}(\xi)\,.
\label{10_CL}
\eea

Expressions (\ref{sup_tau})-(\ref{10_CL}) allow one to construct conserved charges ${\cal P}(\xi)$ and their fluxes by surface integration:
\bea
{\cal P}(\xi) &=& \oint_{\di\Sigma}dx^{D-2}{\cal I}^{01}(\xi)
\,,
\m{charge}\\
\dot{\cal P}(\xi) &=& - \oint_{\di\Sigma}dx^{D-2}{\cal I}^{1}(\xi)
\,.
\m{flux}
\eea
Here and below, $\di\Sigma$ is defined by intersection of a hypersurface defined by $r= \const$ with a lightlike section $\Sigma := v = \const$; `overdot' means $d/dv$.
By the definition, the quantity ${\cal I}^{1}(\xi)$ presents the flux density of the quantity ${\cal P}(\xi)$ through the surface $\di\Sigma$. We choose the minus sign in (\ref{flux}) because in our consideration the flux is opposite to the direction of the coordinate $r$. Formally, this is explained by the antisymmetry ${\cal I}^{01} = -{\cal I}^{10}$ and the definition for the current component ${\cal I}^{1}$ in (\ref{10_CL}).

The quantities (\ref{charge}) and (\ref{flux}) can be interpreted as energy inside $\di\Sigma$ and energy flux through $\di\Sigma$, respectively, if $\xi^\alf = \bar\xi^\alf$ is a timelike  Killing vector. For the background metric (\ref{metric_bar}) $\bar\xi^\alf$ has the general form in all the cases of $\bar f(r)$:
\be
\bar\xi^\alf = \l\{-1, 0 ,0,\ldots,0\r\}; \qquad \bar\xi_\alf = \l\{\bar f(r), - 1 ,0,\ldots,0\r\} \,.
\label{bar_xi}
\ee

In our study we will consider observers freely and radially falling in a geometry of a static black hole that is considered as a background for related dynamic solutions.  For the backgrounds of the form (\ref{metric_bar}) a proper vector for such observers is
\be
\xi^\alf = \l\{-|c|\l(1+\sqrt{1-|c|\bar f}\r)^{-1}, \sqrt{1-|c|\bar f} ,0,\ldots,0\r\}\,.
\label{xi_static}
\ee
Constant of integration $c$ has to satisfy the inequality $1-|c|\bar f \geq 0$ and will be concretized later for each of the cases of black holes with AdS, dS and flat asymptotics.

The light cones for the solutions (\ref{metric}) and (\ref{metric_bar}) are defined as follows
\be
\l.\frac{dv}{dr}\r|_{\rm in}=0;~~ \l.\frac{dv}{dr}\r|_{\rm out}={2}/{f}\,~~~~~~ {\rm and} ~~~~~~  \l.\frac{dv}{dr}\r|_{\rm in}=0;~~ \l.\frac{dv}{dr}\r|_{\rm out}={2}/{\bar f}\,,
\label{b+d_cone}\\
\ee
respectively. By definition, a vector (\ref{xi_static}) is placed inside the background light cone in (\ref{b+d_cone}). However, generally, it is not so for the dynamic light cone in (\ref{b+d_cone}). As a first step, to restrict the observer with (\ref{xi_static}) to a physically permissible one, we derive from (\ref{xi_static}) the expression:
\be
\l.\frac{dv}{dr}\r|_{\rm obs}= -\frac{|c|}{(1-|c|\bar f) + \sqrt{1-|c|\bar f}}\,.
\label{obs_cone}
\ee
Then, its combination with the expressions for dynamic light cone  in (\ref{b+d_cone}) gives the relation
\be
- \l.\frac{|c|}{(1-|c|\bar f) + \sqrt{1-|c|\bar f}}\r|_{r=r_R} = \l.\frac{2}{f}\r|_{r=r_R}\,.
\label{r_R}
\ee
A solution to this equation $r_R$ determines a real observer with a proper vector (\ref{xi_static}) inside the dynamic cone only when $r>r_R$. Note that (\ref{r_R}) has solutions when $f$ becomes negative.

To calculate the components (\ref{01_CL}) and (\ref{10_CL})  of the current  for the observer with (\ref{xi_static})  we derive the component of the superpotential (\ref{sup_tau}):
\be
{\cal I}^{01}(\xi) = -(D-2)\frac{\sqrt{q}p}{2\k}\alf^{2(p-1)}r^{D-2p-1}\Delta f(1-\bar f)^{p-1}\frac{|c|}{1+\sqrt{1-|c|\bar f}} \,.
\label{sup_tau_1}
\ee
This gives the energy density measured by the aforementioned observer:
\bea
{\cal I}^{0}(\xi) &=& -(D-2)\frac{\sqrt{q}p}{2\k}\alf^{2(p-1)}r^{D-2p-1}\Delta f(1-\bar f)^{p-1}\frac{|c|}{1+\sqrt{1-|c|\bar f}} \nonumber\\
&\times & \l[ \frac{D-2p-1}{r} + \frac{f' - \bar f'}{\Delta f}-(p-1)\frac{\bar f'}{1-\bar f}+ \half \frac{|c|\bar f'}{(1-|c|\bar f) + \sqrt{1- |c|\bar f}}\r]\,,
\label{01_CL_1}
\eea
and a related density of the energy flux:
\be
{\cal I}^{1}(\xi) = (D-2)\frac{\sqrt{q}p}{2\k}\alf^{2(p-1)}r^{D-2p-1}(1-\bar f)^{p-1}\frac{|c| \dot f}{1+\sqrt{1-|c|\bar f}} \,.
\label{10_CL_1}
\ee

\section{Black hole solutions. The case $1/\ell^2<0$ }
 \m{BH_AdS}
\setcounter{equation}{0}

Let us begin by considering (\ref{even+}) and (\ref{odd+}) as static solutions setting $m(v) = {\cal C}(v) = 0$. For $1/\ell^2<0$ and even $p$ there are no black hole solutions  \cite{Cai_Ohta_2006}, and we do not consider them. In the case of odd $p$ and $1/\ell^2<0$ the solution (\ref{odd+}) becomes
\be
f(r) = 1 +  \frac{r^2}{\alf^{2-2/p}}\l(\frac{1}{\l|\ell\r|^2} - \frac{ m_0}{r^{D-1}}\r)^{1/p}
\,.
\m{odd_AdS}
\ee
This presents a static black hole with the event horizon $r_+$ satisfying $f(r_+)=0$, or
\be
\frac{r_+^{2p}}{\l|\ell\r|^2} + \alf^{2p-2} =  \frac{ m_0}{r_+^{D-2p-1}}
\,.
\m{horizon_AdS}
\ee
The last presents the unique horizon because $D \ge 2p+2$ and $m_0>0$.
To define a dynamic black hole we assume that the light-like fluid, see (\ref{T_light}) - (\ref{Theta_2}), falls onto the black hole defined by  (\ref{odd_AdS}). Then
the Vaidya-like solution (\ref{odd+}) becomes
\be
f(r,v) = 1 +  \frac{r^2}{\alf^{2-2/p}}\l(\frac{1}{\l|\ell\r|^2} - \frac{ m_0 + m(v)}{r^{D-1}}- \frac{ {\cal C}(v)\Theta(r)}{r^{D-1}}\r)^{1/p}
\,.
\m{odd_AdS_V}
\ee

Solutions (\ref{odd_AdS}) and (\ref{odd_AdS_V}) have the unique AdS asymptotics:
\be
\bar f(r) = 1 +  \frac{r^2}{\alf^{2-2/p}}\frac{1}{\l|\ell\r|^{2/p}}\,
\m{bar_AdS}
\ee
with the AdS radius $\ell_{AdS} = \alf (\l|\ell\r|/\alf)^{1/p}$. We call these objects as asymptotically AdS black holes. The metric (\ref{metric_bar}) with (\ref{bar_AdS}) can be considered as a background one for the asymptotically AdS black holes of both the types (\ref{odd_AdS}) and (\ref{odd_AdS_V}). Besides, in the case of (\ref{odd_AdS_V}), it is just natural to choose the solution with (\ref{odd_AdS}) as  a background one, now marked by $\bar f$. Then one considers falling matter and  metric perturbations with respect to the geometry of the static black hole.

Because the solution with (\ref{odd_AdS_V}) is a dynamic one its dynamic event horizon, $r_+$, is presented by a more complicated one than (\ref{horizon_AdS}) equation:
\be
f\l(r_+(v),v \r) = 2\frac{dr_+(v)}{dv}
\,,
\m{horizon_AdS_V}
\ee
see, for example, \cite{Ashtekar_Krishnan_2004,Booth_2005,Nielsen_2010,Nielsen+_2011}. Besides, for the solution with (\ref{odd_AdS_V}) a so-called apparent horizon, $r_A$, can be defined by
\be
f\l(r_A(v),v \r) = 0
\,,
\m{horizon_A}
\ee
see, for example, \cite{Ashtekar_Krishnan_2004,Booth_2005,Nielsen_2010,Nielsen+_2011}. Note that for black holes with all the aforementioned asymptotics 1) the definitions (\ref{horizon_AdS_V}) and (\ref{horizon_A}) hold, and 2) apparent horizons (\ref{horizon_A}) coincide with event horizons (\ref{horizon_AdS_V}) for static solutions.

\subsection{The energy of the static black hole}
\m{E_AdS_SBH}

Consider the solution (\ref{odd_AdS}) choosing (\ref{bar_AdS}) as a background one, then
\be
\Delta f = f(r) - \bar f(r)= \frac{r^2}{\alf^{2-2/p}}\l[\l(\frac{1}{\l|\ell\r|^2} - \frac{ m_0}{r^{D-1}}\r)^{1/p} - \frac{1}{\l|\ell\r|^{2/p}}\r]
\,.
\m{odd_AdS_Delta}
\ee
Now we calculate the global energy of the black hole (\ref{odd_AdS}) with respect to AdS background (\ref{bar_AdS}). We use the surface integration (\ref{charge}) with the superpotential component (\ref{sup_tau}), where the timelike Killing vector (\ref{bar_xi}) is related to a background spacetime with (\ref{bar_AdS}), the perturbation $\Delta f$ is defined in (\ref{odd_AdS_Delta}) and $r \goto \infty$. Then
\be
\l.{\cal P}(\bar\xi)\r|_{r \goto \infty}  = \lim_{r \goto \infty}\oint_{\di\Sigma}dx^{D-2}{\cal I}^{01}(\bar\xi)= \l(D-2 \r)\frac{m_0}{4G_D}\,.
\m{P_0_AdS}
\ee
The result (\ref{P_0_AdS}) coincides  with the result obtained by us in \cite{Petrov_2019} in the case of a general set of coupling constants $\{\alf_p\}$ under the assumption of existence  of black hole solutions with the AdS asymptotics. The result (\ref{P_0_AdS}) is in an agreement with the earlier results that take various approaches in different theories, see \cite{DT1,DT2,AFrancavigliaR,Okuyama,DerKatzOgushi,Kofinas_Olea_2007,Peng_Liu_2020}.

The result (\ref{P_0_AdS}) presents the global energy. However, formula (\ref{charge}) allows us to construct quasi-local characteristics as well. Let us consider observers at rest with the constant $r=r_0$. This is possible when the observers are outside the event horizon. As a consequence of preserving the spherical symmetry, one finds that their proper vectors coincide with the Killing vector (\ref{bar_xi}).
As a result, one obtains the quasi-local energy for such observers:
\be
\l.{\cal P}(\bar\xi)\r|_{r = r_0} = \oint_{r_0}dx^{D-2}{\cal I}^{01}(\bar\xi) = -\frac{(D-2)}{4G_D}\frac{p}{|\ell|^2}r^{D-1}_0\l[\l(1- \frac{|\ell|^2 m_0}{r^{D-1}_0} \r)^{1/p} -1 \r]
\,.
\m{quasi_local+}
\ee
This is interpreted as the energy under the surface $r=r_0$. Assuming ${|\ell|^2 m_0}/{r^{D-1}_0}\ll 1$, one has
\be
\
\l.{\cal P}(\bar\xi)\r|_{r = r_0} \approx \frac{(D-2)}{4G_D}\l[m_0 + \frac{1}{2!}\l(1- \frac{1}{p}\r)\frac{|\ell|^2 m^2_0}{r^{D-1}_0} + \ldots \r]
\,.
\m{quasi_local_sim}
\ee
This is more than the global energy (\ref{P_0_AdS}). Thus, the energy outside the surface $r=r_0$ has to be negative. Because the exact expression (\ref{quasi_local+}) is monotonic in radial coordinate this claim is true in the exact sense as well. Since $r=r_0$ is arbitrary, the energy density of free gravitational field outside the horizon is negative. This is quite interesting because this claim, in fact, coincides with the analogous results in 4-dimensional Einstein gravity, see, for example, \cite{Petrov+_2017,Misner_Thorn_Wheeler_1973,Brown_York_1993} and many references therein.

\subsection{The energy of the dynamic black hole}
\m{E_AdS_DBH}

Let us turn to the dynamic solution (\ref{odd_AdS_V}) choosing (\ref{bar_AdS}) as a background again, then
\be
\Delta f = f(r,v) - \bar f(r)= \frac{r^2}{\alf^{2-2/p}}\l[\l(\frac{1}{\l|\ell\r|^2} - \frac{ m_0 + m(v)}{r^{D-1}}- \frac{ {\cal C}(v)\Theta(r)}{r^{D-1}}\r)^{1/p} - \frac{1}{\l|\ell\r|^{2/p}}\r]
\,.
\m{odd_AdS_Delta_A}
\ee
Firstly, consider as an example the pure radiation that is the null dust without pressure when ${\cal C}(v)=0$. In this case, in analogy with (\ref{P_0_AdS}), one finds the global energy for the solution (\ref{odd_AdS_V}):
\be
\l.{\cal P}(\bar\xi)\r|_{r \goto \infty} =  \l(D-2 \r)\frac{m_0+m(v)}{4G_D}
\,
\m{P_0_AdS_V}
\ee
at each of the moment $v$. Secondly, consider a more complicated case with presence of pressure $P_r$ when ${\cal C}(v)\neq 0$. A question arises: can the last term in brackets in (\ref{odd_AdS_V}) contribute to the global  energy? We turn to the cases of behavior for $\Theta(r)$ in (\ref{Theta_1}) and (\ref{Theta_2}). Substituting (\ref{Theta_1}) into the integral of the type (\ref{P_0_AdS}) one finds that $\l.{\cal P}(\bar\xi)\r|_{r \goto \infty}$ diverges and therefore has no a physical sense. When $\sigma > -1/(D-2)$ in (\ref{Theta_2}) the quantity $\l.{\cal P}(\bar\xi)\r|_{r \goto \infty}$ diverges as well. Only for
\be
-1\le \sigma < -1/(D-2)\,
\m{sigma}
\ee
the quantity $\Theta(r)$ in (\ref{Theta_2}) can be taken into account. However, it does not contribute  into the global result (\ref{P_0_AdS_V}). Thus, examining the global energy, we exclude (\ref{Theta_1}) from the consideration and restrict $\sig$ by (\ref{sigma}) in (\ref{Theta_2}).

Both of the results (\ref{P_0_AdS}) and (\ref{P_0_AdS_V}) for the global energy have been obtained with respect to AdS background with (\ref{bar_AdS}). However, for the solution (\ref{odd_AdS_V}) it is natural to use the static black hole solution (\ref{odd_AdS})  as a background. Formula of the type (\ref{P_0_AdS}) can be used again.  Only now $\Delta f$ is defined as
\be
\Delta f = \frac{r^2}{\alf^{2-2/p}}\l[\l(\frac{1}{\l|\ell\r|^2} - \frac{ m_0}{r^{D-1}} - \frac{m(v) + {\cal C}(v)\Theta(r)}{r^{D-1}}\r)^{1/p} - \l(\frac{1}{\l|\ell\r|^2} - \frac{ m_0}{r^{D-1}}\r)^{1/p}  \r]\,,
\m{Delta_AdS_V}
\ee
and $\bar f$ in formula (\ref{bar_xi}) for Killing vector has to be taken from (\ref{odd_AdS}).
Then,
\be
\l.{\cal P}(\bar\xi)\r|_{r \goto \infty} =  \l(D-2 \r)\frac{m(v)}{4G_D}
\,.
\m{P_AdS_static}
\ee
The restrictions (\ref{sigma}) hold again and are taken into account.
We stress that the {\em global} energy is additive because (\ref{P_0_AdS_V}) is a sum of  (\ref{P_0_AdS}) and  (\ref{P_AdS_static}).

Now, let us consider quasi-local energy for the solution (\ref{odd_AdS_V}) when the observer is at the rest at $r=r_0$. This is possible when $r_0> r_+(v)$ defined by (\ref{horizon_AdS_V}) on the sections $\Sigma := v = \const$. As a background we consider both (\ref{bar_AdS}) and (\ref{odd_AdS}). Then, in the first case, $\Delta f$ is defined by (\ref{odd_AdS_Delta_A}) at fixed $r=r_0$.
Substituting such $\Delta f$ into (\ref{sup_tau}) with (\ref{bar_xi}) and calculating with the formula of the type (\ref{quasi_local+}) one obtains
\be
\l.{\cal P}(\bar\xi)\r|_{r = r_0} = -\frac{(D-2)}{4G_D}\frac{p}{|\ell|^2}r^{D-1}_0
\l[\l(1 - \frac{|\ell|^2\l[m_0+ m(v) +
{\cal C}(v)\Theta(r_0)\r]}{r_0^{D-1}}\r)^{1/p} -1 \r]
\,.
\m{quasi_local_AB}
\ee
In the case of the background (\ref{odd_AdS}), $\Delta f$ is defined by (\ref{Delta_AdS_V}) at fixed $r=r_0$.
Then, substituting such $\Delta f$ into (\ref{sup_tau}) with (\ref{bar_xi}) and calculating with (\ref{charge}) at $r=r_0$, one obtains
\bea
\l.{\cal P}(\bar\xi)\r|_{r = r_0} &=& -\frac{(D-2)}{4G_D}p\l(\frac{1}{|\ell|^2}- \frac{m_0}{r^{D-1}_0}\r)r^{D-1}_0\nonumber\\
&\times&\l[\l(1 - \frac{ m(v) +
{\cal C}(v)\Theta(r_0)}{r_0^{D-1}}\l(\frac{1}{|\ell|^2}- \frac{m_0}{r^{D-1}_0}\r)^{-1}\r)^{1/p} -1 \r]
\,.
\m{quasi_local_V}
\eea
Observing (\ref{quasi_local_AB}) and (\ref{quasi_local_V}), we conclude: 1) Because $m(v) + {\cal C}(v)\Theta(r)$ is increasing with $v$ the total energy (of gravitational field with radiating matter) outside $r=r_0$ is negative. 2) Pressure determined by ${\cal C}(v)$ contributes to the quasi-local energy. 3) The quasi-local energy is not additive. Indeed, (\ref{quasi_local+}) plus (\ref{quasi_local_V}) is not equal to (\ref{quasi_local_AB}) even when ${\cal C}(v)=0$.

\subsection{The energy flux for the dynamic black hole}
\m{F_AdS_DBH}

Because the solution (\ref{odd_AdS_V}) is a dynamic one it is interesting to derive the rate of change of the energy, i.e. its velocity.
The energy flux is calculated by the standard formula (\ref{flux}). The definition of ${\cal I}^{1}(\bar\xi)$ is given in (\ref{10_CL}) with (\ref{sup_tau}). Keeping in mind all the above approximations at $r\goto\infty$ with the background (\ref{bar_AdS}) as well as with (\ref{odd_AdS}), one obtains for the flux of the global energy
\be
\l.\dot{\cal P}(\bar\xi)\r|_{r \goto \infty} =  \l(D-2 \r)\frac{\dot m(v)}{4G_D}
\,.
\m{P_AdS_velocity}
\ee
Of course, the same result (\ref{P_AdS_velocity}) is obtained by a direct differentiating (\ref{P_0_AdS_V}) or (\ref{P_AdS_static}).

Analogously, we calculate the flux of the quasi-local energy (\ref{quasi_local_AB}):
\be
\l.\dot{\cal P}(\bar\xi)\r|_{r = r_0} = \frac{(D-2)}{4G_D}\l[\dot m(v) + \dot{\cal C}(v)\Theta(r_0) \r]
\l(1 - \frac{|\ell|^2\l[m_0+ m(v) +
{\cal C}(v)\Theta(r_0)\r]}{r_0^{D-1}}\r)^{1/p-1}
\,,
\m{flux_local_AB}
\ee
and of the quasi-local energy (\ref{quasi_local_V}):
\bea
\l.\dot{\cal P}(\bar\xi)\r|_{r = r_0} &=& \frac{(D-2)}{4G_D}\l[\dot m(v) + \dot{\cal C}(v)\Theta(r_0) \r]\nonumber\\&\times&
\l(1 - \frac{ m(v) +
{\cal C}(v)\Theta(r_0)}{r_0^{D-1}}\l(\frac{1}{|\ell|^2}- \frac{m_0}{r^{D-1}_0}\r)^{-1}\r)^{1/p-1}
\m{flux_local_V}
\eea
with the similar explanations as for (\ref{P_AdS_velocity}).

\subsection{A freely falling observer}
\m{AdS_freely}

In the case of a dynamic black hole it is interesting to derive local characteristics expressed by the components (\ref{01_CL}) and (\ref{10_CL}) of the current ${\cal I}^\alf(\xi)$ for a moving observer. For the role of observer we choose a freely falling particle on the background (\ref{odd_AdS}) with a proper vector of the general form (\ref{xi_static}).  As integration constant we choose $|c| = 1$.
Then, from (\ref{xi_static}), it follows that the radial falling starts from the rest at $r_* = \l(m_0 |\ell|^2\r)^{1/(D-1)}$.

In the case (\ref{odd_AdS}) the proper vector of the observer (\ref{xi_static}) is inside the dynamic light cone in (\ref{b+d_cone}) when $f\geq 0$ that is
 outside the apparent horizon, $r \geq r_A(v)$, defined by (\ref{horizon_A}). Besides, the event horizon defined by (\ref{horizon_AdS_V}) is outside the apparent horizon defined by (\ref{horizon_A}). In addition, because $m(v) + {\cal C}(v)\Theta(r)$ is increasing with $v$, one easily shows ${dr_+(v)}/{dv}>0$. This leads to $f(r_+(v),v) > f(r_A(v))$ that holds only when $r_+> r_A$. Thus, the proper vector of the observer  (\ref{xi_static}) is inside the dynamic light cone up to the dynamic event horizon as well. Finally, for $0<r<r_A$ there is $r=r_R$ defined by (\ref{r_R})
that restricts a  possibility to consider such a freely falling observer.

Under the aforementioned conditions we are in a position to present the local energetic characteristics ${\cal I}^\alf(\xi)$ for the observer with the proper vector (\ref{xi_static}). Because the expressions are very cumbersome we have moved them to Appendix \ref{case_AdS}.

\section{Black hole solutions. The case $1/\ell^2>0$ }
 \m{BH_dS}
\setcounter{equation}{0}

Let us consider (\ref{even+}) and (\ref{odd+}) for $1/\ell^2>0$ as static solutions setting $m(v) = {\cal C}(v) = 0$. For even $p$ the solution (\ref{even+}) with the `plus' branch can be interpreted as a naked singularity \cite{Cai_Ohta_2006}, and we do not consider it. Then both (\ref{even+}) for the even $p$  with the `minus' branch and the  solution (\ref{odd+}) for the odd $p$ are combined as
\be
f(r) = 1 - \frac{r^2}{\alf^{2-2/p}}\l(\frac{1}{\ell^2} + \frac{ m_0}{r^{D-1}}\r)^{1/p}
\,.
\m{dS_1}
\ee
Let us find conditions when (\ref{dS_1}) presents a black hole. Vanishing $m_0$ in (\ref{dS_1}) gives
\be
\bar f(r) = 1 -  \frac{r^2}{\alf^{2-2/p}}\frac{1}{\ell^{2/p}}\,
\m{dS_2}
\ee
that represents dS space with the dS radius $\ell_{dS} = \alf (\ell/\alf)^{1/p}$. The other role of $\ell_{dS}$ in dS space is a cosmological horizon denoted as $r_c$. The presence of $r_c$ excludes the limit $r\goto \infty$, unlike the AdS case. Returning to (\ref{dS_1}), we define the event horizon by $f(r_+)=0$ that is
\be
 \alf^{2p-2} = \frac{r_{+}^{2p}}{\ell^2} + \frac{ m_0}{r_{+}^{D-2p-1}}
\,.
\m{dS_3}
\ee
Let us analyse it.
If $m_0$ is equal to a critical value $m_{\rm cr}$:
\be
 m_{\rm cr} = 2p\,\alf^{(D-1)(2-2/p)}\,\ell^{(D-2p-1)/p}\,(D-2p-1)^{(D-2p-1)/2p}\,(D-1)^{-(D-1)/2p}
\,,
\m{dS_4}
\ee
there is one solution to the equation  (\ref{dS_3}). It is a Nariai-like solution with the exotic properties \cite{Nariai_1950,Nariai_1951}. When $m_0 > m_{\rm cr}$ there are no solutions to the equation  (\ref{dS_3}) at all, and (\ref{dS_1}) presents a naked singularity. Thus, we consider the case (\ref{dS_1}) with $m_0< m_{\rm cr}$ only when there are two solutions to the equation (\ref{dS_3}): $r_+$ and $r_c$. In this case a black hole solution exists, while the second solution $r_c > r_+$ represents a cosmological horizon.

For the dynamic black hole we assume that the light-like fluid, see (\ref{T_light}) - (\ref{Theta_2}), falls onto the black hole presented by (\ref{dS_1}). In this case, the solution (\ref{dS_1}) becomes
\be
f(r,v) = 1 -  \frac{r^2}{\alf^{2-2/p}}\l(\frac{1}{\ell^2} + \frac{ m_0 + m(v)}{r^{D-1}}+ \frac{ {\cal C}(v)\Theta(r)}{r^{D-1}}\r)^{1/p}
\,,
\m{dS_5}
\ee
see (\ref{even+}) and (\ref{odd+}). Then we require that the equation $f(r_+,v_0)=0$, that is
\be
 \alf^{2p-2} = \frac{r_{+}^{2p}}{\l|\ell\r|^2} + \frac{ m_0+ m(v_0)+{\cal C}(v_0)\Theta(r_{+}) }{r_{+}^{D-2p-1}}\,,
\m{dS_6}
\ee
has to have two different solutions: $r_+$ and $r_c$. Since we are restricted by monotonic dynamic, a black hole exists in interval $0\leq v\leq v_0$ as well.

Objects related to both the  solutions (\ref{dS_1}) and (\ref{dS_5}) are called asymptotically dS black holes. One can choose the dS background (\ref{dS_2}) for these black holes. Besides, it is natural to choose the geometry of the static black hole (\ref{dS_1}) as a background one for the dynamic solution with (\ref{dS_5}).

\subsection{Quasi-local conserved quantities}
\m{QL_dS}

Because for asymptotically dS black holes there is no limit $r \goto \infty$, one can only calculate quasi-local quantities. We consider observers at the rest with the constant $r=r_0$ when $r_+<r_0<r_c$. Consider the static solution (\ref{dS_1}), choosing (\ref{dS_2}) as a background, then,
\be
\Delta f = f(r_0) - \bar f(r_0)= - \frac{r_0^2}{\alf^{2-2/p}}\l[\l(\frac{1}{\ell^2} + \frac{ m_0}{r_0^{D-1}}\r)^{1/p} - \frac{1}{\ell^{2/p}}\r]
\,.
\m{dS_9}
\ee
Once again, to find the quasi-local energy we use the surface integration (\ref{charge}). Now the superpotential component (\ref{sup_tau}) is calculated for (\ref{dS_9}) with the timelike Killing vector (\ref{bar_xi})  related to a background (\ref{dS_2}). Finally one gets
\be
\
\l.{\cal P}(\bar\xi)\r|_{r = r_0} = \frac{(D-2)}{4G_D}\frac{p}{\ell^2}r^{D-1}_0\l[\l(1+ \frac{\ell^2 m_0}{r^{D-1}_0} \r)^{1/p} -1 \r]
\,.
\m{dS_11}
\ee
This is interpreted as energy under the surface $r=r_0$. The value of this expression is monotonically increasing with increasing $r_0$. Thus, the energy density of free gravitational field can be seen to be positive. This quite differs from the conclusion in the case of the asymptotically AdS black holes. We would want to stress that asymptotically dS black holes frequently have unusual properties. For example, in \cite{dS_1,dS_2} authors have proved that, in models where static black holes are placed in dS space, the full mass is less than the mass of non-singular dS space without a black hole.

Choosing (\ref{dS_2}) as a background for the dynamic solution (\ref{dS_5}), one has
\be
\Delta f =  - \frac{r_0^2}{\alf^{2-2/p}}\l[\l(\frac{1}{\ell^2} + \frac{ m_0 + m(v)}{r_0^{D-1}}+ \frac{ {\cal C}(v)\Theta(r_0)}{r_0^{D-1}}\r)^{1/p} - \frac{1}{\ell^{2/p}}\r]
\,.
\m{dS_12}
\ee
The quasi-local energy becomes:
\be
\l.{\cal P}(\bar\xi)\r|_{r = r_0} = \frac{(D-2)}{4G_D}\frac{p}{\ell^2}r^{D-1}_0
\l[\l(1 + \frac{\ell^2\l[m_0+ m(v) +
{\cal C}(v)\Theta(r_0)\r]}{r_0^{D-1}}\r)^{1/p} -1 \r]
\,.
\m{dS_13}
\ee
Due to the restriction in monotonic evolution we have the interpretation of (\ref{dS_13}) as follows.  The density of the total energy of the gravitational field together with the null fluid is  positive. Because in the dS case there is no well defined limit at $r\goto\infty$ one has no possibility to incorporate the restriction like (\ref{sigma}) for $\sigma$.
Thus, for the definition of $\Theta(r)$  in quasi-local expression we can use both of possibilities (\ref{Theta_1}) and (\ref{Theta_2}).

In the case of the dynamic solution (\ref{dS_5}) with the static background (\ref{dS_1}) we derive
\be
\Delta f = - \frac{r^2_0}{\alf^{2-2/p}}\l[\l(\frac{1}{\ell^2} + \frac{ m_0 }{r_0^{D-1}}+ \frac{ { m(v)+\cal C}(v)\Theta(r_0)}{r_0^{D-1}}\r)^{1/p} - \l(\frac{1}{\ell^2} + \frac{ m_0}{r^{D-1}_0}\r)^{1/p}  \r]\,.
\m{dS_14}
\ee
Then, the related expression for the quasi-local energy is
\bea
\l.{\cal P}(\bar\xi)\r|_{r = r_0} &=& \frac{(D-2)}{4G_D}p\l(\frac{1}{\ell^2}+ \frac{m_0}{r^{D-1}_0}\r)r^{D-1}_0\nonumber\\
&\times&\l[\l(1 + \frac{ m(v) +
{\cal C}(v)\Theta(r_0)}{r_0^{D-1}}\l(\frac{1}{\ell^2}+ \frac{m_0}{r^{D-1}_0}\r)^{-1}\r)^{1/p} -1 \r]
\,.
\m{dS_15}
\eea
Qualitatively the interpretation of (\ref{dS_15}) is similar to the interpretation of (\ref{dS_13}). In addition, we note that in the case of asymptotically dS black holes the quasi-local energy is not additive.

Finally, in analogy to (\ref{flux_local_AB}) and (\ref{flux_local_V}), we calculate the flux of the quasi-local energy (\ref{dS_13}),
\bea
\l.\dot{\cal P}(\bar\xi)\r|_{r = r_0} &=& \frac{(D-2)}{4G_D}\l[\dot m(v) + \dot{\cal C}(v)\Theta(r_0) \r]\nonumber\\&\times&
\l(1 + \frac{\ell^2\l[m_0+ m(v) +
{\cal C}(v)\Theta(r_0)\r]}{r_0^{D-1}}\r)^{1/p-1}
\,,
\m{dS_16}
\eea
 and the flux of the quasi-local energy (\ref{dS_15}),
\bea
\l.\dot{\cal P}(\bar\xi)\r|_{r = r_0} &=& \frac{(D-2)}{4G_D}\l[\dot m(v) + \dot{\cal C}(v)\Theta(r_0) \r]\nonumber\\&\times&
\l(1 + \frac{ m(v) +
{\cal C}(v)\Theta(r_0)}{r_0^{D-1}}\l(\frac{1}{\ell^2}- \frac{m_0}{r^{D-1}_0}\r)^{-1}\r)^{1/p-1}\,.
\m{dS_17}
\eea

\subsection{A freely falling observer}
\m{FFO_dS}

In the present subsection we construct local characteristics expressed by components of the current ${\cal I}^\alf(\xi)$ in the case of the dynamic asymptotically dS black holes (\ref{dS_5}).
As before, for the role of the observer we choose a freely and radially falling particle, now on the background of (\ref{dS_1}). Its proper vector again has the general form (\ref{xi_static}), however, we now choose the constant of integration as
\be
|c| = \frac{1}{1- \frac{r_*}{\alf^{2-2/p}\ell^{2/p}}\l(1+\frac{2p}{D-2p-1} \r)^{1/p}};\qquad r_* \equiv \l[\frac{1}{2p}(D-2p-1) m_0 \ell^2 \r]^{1/(D-1)}\,.
\m{const_2}
\ee
The observer falls from the rest at $r_*$ that defines a maximum of $\bar f$ when $r_+< r_* <r_c$.

Because (\ref{obs_cone}) is negative, the observer is inside the dynamic light cone in (\ref{b+d_cone}) when $f\geq 0$ for (\ref{dS_5}). This takes place
outside the apparent horizon, $r \geq r_A(v)$, defined by (\ref{horizon_A}). For (\ref{dS_5}) in the region $0<r<r_A$ there exists also $r=r_R$ defined by (\ref{r_R})
that restricts the possibility to use such a freely falling observer.

In the asymptotically dS black hole case, it is not simple to define the relation between the event horizon defined by (\ref{horizon_AdS_V}) and the apparent horizon defined by (\ref{horizon_A}). On the one hand, keeping in mind the requirement of monotonic behavior in $v$, one finds ${dr_+(v)}/{dv}>0$ leading to the inequality $f(r_+(v),v) > f(r_A(v))$. On the other hand, a qualitative analysis shows that the last inequality can be derived both when $r_+>r_A$ and when $r_+<r_A$ in various regions and in various time intervals. More concrete conclusions can be made using in depth numerical calculations that are outside the scope of the present paper.

Finally, keeping in mind the above conditions and restrictions, we can present the local energetic characteristics ${\cal I}^\alf(\xi)$ for the freely falling observer with the proper vector (\ref{xi_static}) in the dS case. These are presented in Appendix \ref{case_dS}.

\section{Black hole solutions. The case $1/\ell^2=0$ }
 \m{BH_flat}
\setcounter{equation}{0}

In the case $1/\ell^2=0$, restricting ourselves to static black hole solutions (without naked singularities), we unite the solution (\ref{odd+}) and `minus' branch of (\ref{even+}) as
\be
f(r) = 1 - \frac{r^2}{\alf^{2-2/p}} \l(\frac{m_0}{r^{D-1}} \r)^{1/p}\,.
\m{F_1}
\ee
The event horizon is defined as usual, $f(r_+) = 0$. This gives
\be
r_+ = \l(\frac{m_0}{\alf^{2p-2}} \r)^{1/(D-2p-1)}\,.
\m{F_2}
\ee
To define a dynamic black hole let us turn to the solution (\ref{even+}), with the `minus' branch only, and to (\ref{odd+}). Then the solution (\ref{F_1}) is generalized as
\be
f(r,v) = 1 - \frac{r^2}{\alf^{2-2/p}} \l(\frac{m_0  +m(v) +{\cal C}(v)\Theta(r)}{r^{D-1}} \r)^{1/p}\,.
\m{F_3}
\ee
Both (\ref{F_1}) and (\ref{F_3}) have  flat asymptotics, and we call the objects asymptotically flat black holes.

\subsection{The energy of the static black hole}
\m{E_dS_SBH}

The flat background is the most appropriate one for the solutions with flat asymptotics. However, the direct application of the field-theoretical approach is problematic. Indeed, we have to state $\bar R^{\alf\beta}_{\gamma\delta} = 0$ in (\ref{Riemannian_bar}), and, consequently, to set $\bar f=1$. As a result ${\bm \omega}^{\rho\lam |\mu\nu} = 0$ when $p\neq 1$, see (\ref{omega++})-(\ref{www}). The exception is only for the Einstein theory, for which $p=1$. Thus, the suprpotential ${\cal I}^{\mu\nu}(\xi)$ in (\ref{Super-l}) vanishes totally, and it is undetermined.

This is a well known problem in the linearized approaches in constructing conserved quantities when they are applied to the degenerated variants of the Lovelock theory, see \cite{Rodrigo_2017} and references therein, or to the quadratic gravity, see, for example,  \cite{DT1,DT2,Rodrigo+_2020}. In this case the linearized version of equations, see the left hand side of the equations (\ref{GL-q}), disappears. This situation induces the degeneracy of the right hand side of the equations (\ref{GL-q}) and, consequently, the degeneracy  of the conserved quantities in the field-theoretical approach.

Considering the pure Lovelock gravity with $\alf_0=0$ and flat backgrounds $\bar R^{\alf\beta}_{\gamma\delta} = 0$ one finds that the linearized version of the equations, being proportional to $\bar R^{\alf\beta}_{\gamma\delta}$, disappears as well. Moreover, this disappearance takes a place for all variants of the Lovelock gravity (not only for pure one) with the absence of the Einstein-Hilbert action when $\alf_0=0$ and $\bar R^{\alf\beta}_{\gamma\delta} = 0$. The problem can be eliminated by adding the Einstein-Hilbert action.\footnote{One has to note that the methods of constructing conserved charges without backgrounds could resolve the problem of degeneracy. Thus, applying the counterterm technique, see \cite{Olea_2005,Olea_2007,Miskovic_Olea_2007} this problem has been resolved in \cite{Rodrigo_2017,Rodrigo+_2020}.}

Here, we resolve this problem in the pure Lovelock gravity in the framework of the field-theoretical approach by including auxiliary structures.
Let us consider
\be
f(r) = 1 - \frac{r^2}{\alf^{2-2/p}} \l(\frac{m_{au} + m_0}{r^{D-1}} \r)^{1/p}\,
\m{F_4}
\ee
instead of (\ref{F_1}) and choose as a background an auxiliary solution
\be
\bar f(r) = 1 - \frac{r^2}{\alf^{2-2/p}} \l(\frac{m_{au}}{r^{D-1}} \r)^{1/p}\,
\m{F_5}
\ee
with constant $m_{au} >0$. Then, the related auxiliary background curvature tensor $\bar R^{\alf\beta}_{\gamma\delta}(au) \neq 0$ and the problem of degeneracy is fixed. The metric perturbation is defined as
\be
\Delta f(r) = - \frac{r^2}{\alf^{2-2/p}} \l[\l(\frac{m_{au}+ m_0}{r^{D-1}} \r)^{1/p} -  \l(\frac{m_{au}}{r^{D-1}} \r)^{1/p}\r] \,,
\m{F_6}
\ee
and a resulting expression for the superpotential component (\ref{sup_tau}) with the Killing vector related to (\ref{F_5}) acquires the form:
\be
{\cal I}^{01}(\bar \xi) = (D-2) \frac{p\sqrt{q}}{2\k} m_{au} \l[\l(1+ \frac{ m_0}{m_{au}} \r)^{1/p} -  1\r] \,.
\m{F_7}
\ee
Because this expression does not depend on $r$ its surface integration by (\ref{charge}) gives the charge
\be
{\cal P}(\bar \xi) = (D-2) \frac{p}{4G_D} m_{au} \l[\l(1+ \frac{ m_0}{m_{au}} \r)^{1/p} -  1\r] \,,
\m{F_8}
\ee
 which is valid both for arbitrary finite $r_0$ and for $r\goto\infty$. Thus, the energy is fully concentrated inside the sphere with $r^{au}_+$, which is the horizon related to the auxiliary solution (\ref{F_5}):
\be
r_+^{au} = \l(\frac{m_{au}}{\alf^{2p-2}} \r)^{1/(D-2p-1)}\,.
\m{F_9}
\ee
Indeed, a possibility for the Killing vector (\ref{bar_xi}) related  to the auxiliary solution (\ref{F_5}) to be a timelike vector is restricted by $r > r_+^{au}$.

The expression (\ref{F_8}) contains the auxiliary parameter $m_{au}$ that has to be excluded from the final expression. This is achieved by the limit $m_{au} \goto \infty$. Then (\ref{F_8}) goes to the acceptable result, like (\ref{P_0_AdS}),
\be
\l.{\cal P}(\bar \xi)\r|_{r \goto \infty} = (D-2) \frac{m_0}{4G_D} \,.
\m{F_10}
\ee
This result can be interpreted as global energy only because the Killing vector has to be outside the event horizon when $r_+^{au} \goto \infty$. Finally, because the component $\bar\xi_1$ of the Killing vector (\ref{bar_xi}) related to (\ref{F_5}) coincides with $\bar\xi_1$ related to (\ref{F_1}) the result (\ref{F_10}) can be interpreted as an acceptable one for the global energy of the black hole defined by (\ref{F_1}).

It seems that (\ref{F_10}) could be obtained with a limit $1/\ell^2 \goto 0$ in (\ref{odd_AdS}) in the framework of the study in section \ref{BH_AdS}. However, it is completely appropriate since the solution with (\ref{odd_AdS}) is valid for only odd $p$, whereas (\ref{F_1}) is defined for all $p$.

\subsection{Dynamic black holes}
\m{E_dS_DBH}

In the case of the dynamic black hole with (\ref{F_3}), it is impossible to use a flat background in the framework of the field-theoretical approach as well. In this case, we use the auxiliary solution (\ref{F_5}) as a background again. The expression for the superpotential (\ref{sup_tau}) with the Killing vector related to (\ref{F_5}) acquires the form:
\be
{\cal I}^{01}(\bar \xi) = (D-2) \frac{p\sqrt{q}}{2\k} m_{au} \l[\l(1+ \frac{ m_0+ m(v) + {\cal C}(v)\Theta(r)}{m_{au}} \r)^{1/p} -  1\r] \,.
\m{F_11}
\ee
Then, integration by (\ref{charge}) gives
\be
\l.{\cal P}(\bar \xi)\r|_{r\goto\infty} = (D-2) \frac{m_0+ m(v)}{4G_D} \,
\m{F_12}
\ee
at each of the moment $v$, under the limit $m_{au}\goto\infty$, and taking into account the definition of $\Theta(r)$ in (\ref{Theta_1}) and (\ref{Theta_2}) restricted by (\ref{sigma}).

Otherwise, the field-theoretical method allows us to study the dynamic black hole (\ref{F_3}) on the background of a static solution (\ref{F_1}), denoting it now as $\bar f$. Let us derive
\be
\Delta f = - \frac{r^2}{\alf^{2-2/p}} \l[\l(\frac{m_0+ m(v) + {\cal C}(v)\Theta(r)}{r^{D-1}} \r)^{1/p} -  \l(\frac{m_0}{r^{D-1}} \r)^{1/p}\r] \,.
\m{F_13}
\ee
Keeping in mind the restriction  (\ref{sigma}) for the definition of  $\Theta(r)$ in  (\ref{Theta_1}) and (\ref{Theta_2}), we obtain for the global energy:
\be
\l.{\cal P}(\bar \xi)\r|_{r\goto\infty} = (D-2) \frac{p}{4G_D} m_0 \l[\l(1+ \frac{ m(v)}{m_0} \r)^{1/p} -  1\r] \,
\m{F_14}
\ee
where the radial pressure does not contribute. For the quasi-local energy we have
\be
\l.{\cal P}(\bar \xi)\r|_{r=r_0} = (D-2) \frac{p}{4G_D} m_0 \l[\l(1+ \frac{ m(v)+ {\cal C}(v)\Theta(r_0)}{m_0} \r)^{1/p} -  1\r] \,.
\m{F_15}
\ee

Let us make some remarks. First, unlike the AdS case, the global energy is not additive  with respect to the background of a static black hole. Indeed, the sum of (\ref{F_10}) and (\ref{F_14}) does not give (\ref{F_12}). Second, in absence  of pressure, ${\cal C}(v)=0$, the quasi-local expression (\ref{F_15}) coincides with the global one (\ref{F_14}). Thus, there is no energy density outside the sphere with radius $r_+$ defined by (\ref{F_2}). The situation could be interpreted in this way when the positive energy density defined by $m(v)$ is compensated by the negative energy density of the free gravitational field. Third, comparing (\ref{F_15}) with (\ref{F_14}) one finds that the energy density outside the sphere with radius $r_0$ is negative. It is quite interesting because this conclusion follows due to the presence in (\ref{F_15}) of pressure only.

At last, we calculate the flux for the global energy by rules of section \ref{BH_AdS}. In the case of the flat background it is
\be
\l.\dot{\cal P}(\bar \xi)\r|_{r\goto \infty} = (D-2) \frac{\dot m(v)}{4G_D}   \,,
\m{F_16+}
\ee
whereas in the case of the background of the static black hole it is
\be
\l.\dot{\cal P}(\bar \xi)\r|_{r\goto \infty} = (D-2) \frac{\dot m(v)}{4G_D} \l(1+ \frac{ m(v)}{m_0} \r)^{1/p-1}  \,.
\m{F_16}
\ee
The flux for the quasi-local energy is
\be
\l.\dot{\cal P}(\bar \xi)\r|_{r=r_0} =  (D-2) \frac{\dot m(v)+ \dot{\cal C}(v)\Theta(r_0)}{4G_D} \l(1+ \frac{ m(v)+ {\cal C}(v)\Theta(r_0)}{m_0} \r)^{1/p-1} \,.
\m{F_17}
\ee

\subsection{A freely falling observer}
\m{FFO_F}

For dynamic black holes (\ref{F_3}) let us derive local characteristics expressed by components of the current ${\cal I}^\alf(\xi)$ in (\ref{01_CL}) and (\ref{10_CL}) with the background metric (\ref{F_1}). For the role of the observer we choose a freely falling particle on the background of (\ref{F_1}) with a proper vector of the general form (\ref{xi_static}). Like in the AdS case, we choose the integration constant as $|c| = 1$. Now the form (\ref{xi_static}) corresponds to the radial falling from the rest at infinity $r_* = \infty$.

For (\ref{F_1}) the observer with (\ref{xi_static}) is inside the dynamic light cone (\ref{b+d_cone}) when for (\ref{F_3}) $f\geq 0$. It takes place outside the apparent horizon, $r \geq r_A(v)$, defined by (\ref{horizon_A}). For the solution (\ref{F_3}), like in the AdS case,  the event horizon defined by (\ref{horizon_AdS_V}) is outside the apparent horizon defined by (\ref{horizon_A}). Indeed, keeping in mind the monotonic evolution, one finds ${dr_+(v)}/{dv}>0$. This leads to the inequality $f(r_+(v),v) > f(r_A(v))$ that uniquely holds when $r_+> r_A$ only. Thus, the observer with (\ref{xi_static}) is inside the dynamic light cone at the dynamic event horizon. Finally, for $0<r<r_A$ there is $r=r_R$ defined by (\ref{r_R})
which restricts the possibility to use the freely falling observer on a background geometry.

Thus, under the aforementioned conditions we can present the local energetic characteristics for the freely falling observer and do this in Appendix \ref{case_flat}.

 \section{Results and discussion}
 \m{Discussion}
\setcounter{equation}{0}

In the present paper, we apply formalism developed in \cite{Petrov_2019} and adopted it to the case of {\em  pure} Lovelock gravity to calculate conserved quantities for both static \cite{Cai_Ohta_2006} and dynamic of the Vaidya type \cite{Cai+_2008} black holes  with AdS, dS and flat asymptotics. In different approaches, conserved charges for static black holes have been found already, see, for example, \cite{Peng_Liu_2020} and references therein. However, to the best of our knowledge, a construction of conserved charges, fluxes, energy densities, energy flux densities for dynamic black holes in complex in pure Lovelock gravity has been absent in literature. We have tried to close this gap, at least partially.

Below we list the results and systematize them:

\noindent {\em I. Static black holes. AdS, dS and flat relative backgrounds:}

\bit
\item[1)] Global energy for AdS (\ref{P_0_AdS}) and flat (\ref{F_10}) black holes.
\item[2)] Quasi-local energy for AdS (\ref{quasi_local+}) and dS (\ref{dS_11}) black holes.

\eit

\noindent {\em II. Dynamic black holes. AdS, dS and flat relative backgrounds:}

\bit
\item[1)] Global energy for AdS (\ref{P_0_AdS_V}) and flat (\ref{F_12}) black holes.
\item[2)] Quasi-local energy for AdS (\ref{quasi_local_AB}) and dS (\ref{dS_13}) black holes.
\item[3)] The flux of global energy for AdS  (\ref{P_AdS_velocity}) and flat (\ref{F_16+}) black holes.
\item[4)] The flux of quasi-local energy for AdS (\ref{flux_local_AB}) and dS (\ref{dS_16}) black holes.
\eit

\noindent {\em III. Dynamic black holes. Backgrounds of related static black holes:}

\bit
\item[1)] Global energy for AdS (\ref{P_AdS_static}) and flat (\ref{F_14}) black holes.
\item[2)] Quasi-local energy for AdS (\ref{quasi_local_V}), dS (\ref{dS_15}) and flat (\ref{F_15}) black holes.
\item[3)] The flux of global energy for AdS  (\ref{P_AdS_velocity}) and flat (\ref{F_16}) black holes.
\item[4)] The flux of quasi-local energy for AdS (\ref{flux_local_V}), dS (\ref{dS_17}) and flat (\ref{F_17}) black holes.
\item[5)] Energy densities for AdS (\ref{I_0_AdS}), dS (\ref{I_0_dS}) and flat (\ref{I_0_flat}) black holes and energy flux densities for AdS (\ref{I_1_AdS}), dS (\ref{I_1_dS}) and flat (\ref{I_1_flat}) black holes in the frame of a freely and radially falling observer .
\eit
Let us now make several comments and discuss the listed results. First of all, we recall that the charges for static black holes are conserved in advanced time $v$, the charges and local quantities for dynamic black holes are defined for each fixed $v$.

Concerning global charges, they are defined for asymptotically AdS and asymptotically flat black holes on AdS and flat backgrounds, respectively, as well as on backgrounds of related static black holes. 1) The requirement of finiteness of global charges restricts $\sigma$ in the energy-momentum (\ref{T_light}) by (\ref{sigma}) instead of (\ref{Theta_1}) and (\ref{Theta_2}). Then, by the correction (\ref{sigma}), the pressure defined by ${\cal C}(v)$ does not contribute to the global charges. 2) For asymptotically AdS black holes the global charges are additive, see a comparison between (\ref{P_0_AdS}), (\ref{P_0_AdS_V}) and (\ref{P_AdS_static}). On the other hand, for asymptotically flat black holes, the global charges are not additive. The formal reason is that on {\em flat backgrounds} the field-theoretical formalism does not work regularly. 3) There is no a possibility to calculate global charges for asymptotically dS black holes. As a result, there are no new restrictions for (\ref{Theta_1}) and (\ref{Theta_2}).

Concerning quasi-local expressions, 1) from formulae (\ref{quasi_local+}) and (\ref{dS_11}) for static black holes one concludes that the energy density of free gravitational field is negative for the asymptotically AdS and positive for the asymptotically dS black holes. 2) Expressions for dynamic black holes (\ref{quasi_local_AB}) and (\ref{dS_13}) on AdS and dS backgrounds as well as (\ref{quasi_local_V}) and (\ref{dS_15}) on backgrounds of static black holes, taking into account the monotonic evolution, lead to conclusion that the energy density of gravitational field together with null fluid is negative for the asymptotically AdS black holes and is positive for the asymptotically dS black holes. 3) Considering asymptotically flat black holes, it is not possible to define quasi-local charges on flat backgrounds. However, quasi-local energy (\ref{F_15}) on the background of static asymptotically flat black holes allows us to conclude that the density of the total energy is negative. It is quite interesting that it is defined by the pressure only! 4) All the aforementioned quasi-local charges are not additive and depend on the pressure term.

It is well known that gravitational energy in metric theories of gravity is not localized, see, for example, the textbook \cite{Misner_Thorn_Wheeler_1973} and the recent review \cite{Petrov_Pitts_2019}.
In this context, the quasi-local definitions are quite successful (see the nice review \cite{Szabados_2009}). Among these the Brown-York quasilocal energy in general relativity is a better known one \cite{Brown_York_1993}. It has been generalized to the pure Lovelock gravity \cite{Chakraborty_Dadhich_2012} recently. However, this approach has been developed for static black holes. In our opinion, the Brown-York method can, at least, be generalized to dynamic black holes considered here. Indeed, the main geometrical object of the Brown-York formalism is a $(D-2)$-sphere $\di\Sigma$ embedded into a physical spacetime as well as into a reference flat spacetime. In the original approach it is defined as $\di\Sigma := t=\const;~r=\const$. It doesn't seem to be impossible to redefine it as $\di\Sigma := v=\const;~r=\const$ and apply to studying dynamic black holes. Then a quasi-local energy for the dynamic black hole has to be defined for each fixed $v$. This is an interesting topic for future study.

Continuing discussion of quasi-local quantities, one has to note that conserved charges for dynamic Vaidya type black holes already have been constructed basing on the  Misner-Sharp quasi-local mass prescription, see \cite{Misner_Sharp_1964} and the generalization in \cite{Maeda_2006}. Thus, in \cite{Cai+_2008} the  Misner-Sharp quasi-local mass has been constructed in the framework of the pure Lovelock gravity; in \cite{Maeda_Nozawa_2008,Nozawa_Maeda_2008} it is constructed and used in the framework of the Einstein-Gauss-Bonnet gravity. In \cite{Maeda+_2011} the  Misner-Sharp quasi-local mass is considered for studying the Vaidya solution in the Lovelock gravity of general type among the other modified theories. This quasi-local quantity has been used, for example, for outlining thermodynamic characteristics at the apparent/trapping horizons of the Vaidya black holes; for stating monotonicity and positivity.

There is a principal difference between the field-theoretical method in constructing conserved quantities (including quasi-local ones) and the Misner-Sharp quasi-local definition of mass. The first method was developed as the Lagrangian based formalism for description of perturbations on fixed backgrounds. A concrete conserved quantities for perturbations can be connected with Killing vectors of backgrounds, with proper vectors of observers connected with backgrounds, etc. Whereas, the Misner-Sharp quasi-local mass definition is not based on any backgrounds; it can be interpreted as a charge connected with the Kodama vector (the last is related to the full geometry without backgrounds). Therefore, it is not surprisingly that the Misner-Sharp quasi-local mass calculated in \cite{Cai+_2008} at fixed $r$ differs from the quasi-local charges obtained here.

However both of the approaches are quite meaningful leading to the standard (all acceptable) global quantities at infinity. Applications for each of the approaches, as a rule, differ. For example, the field-theoretical method cannot be used for describing thermodynamic characteristics at various horizons of dynamic black holes; on the other hand, one cannot apply the Misner-Sharp quasi-local method to present local characteristics for a freely falling observer. Because the Misner-Sharp quasi-local method suggests a construction of conserved charges without backgrounds there are no problem connected with a degeneracy discussed in subsection \ref{E_dS_SBH}. It is an analogy with the counterterm technique \cite{Olea_2005,Olea_2007,Miskovic_Olea_2007}.

Returning to our results related to dynamic black holes, they can be considered as perturbed objects with respect to related static black holes, where perturbations are presented by a light-like fluid and metric perturbations. Freely falling observers measure energetic characteristics of these perturbations. Such observers are physically permissible in regions including apparent horizons of dynamics solutions. Our local expressions are exact, and, of course, can be simplified in the case of infinitesimal perturbations on backgrounds of static black holes if necessary.

\bigskip

\noindent {\bf Acknowledgments}
The author thanks Naresh Dadhich for the idea to apply the field-theoretical formalism for studying solutions in pure Lovelock gravity; he thanks Ricardo Troncoso and Krishnakanta Bhattacharya for discussions and explanation of their works; he thanks also Deepak Baskaran for improving English. The author especially is grateful to anonymous referees for the important comments and recommendations that helped to improve the text significantly. This research has been supported by the Interdisciplinary Scientific and Educational School of Moscow State University ``Fundamental and Applied Space Research''.

\appendix

\section{Local characteristics for freely falling observers}

\subsection{The case of asymptotically AdS black holes}
\m{case_AdS}

Under the conditions given in subsection \ref{AdS_freely} we have calculated the local energetic characteristics ${\cal I}^\alf(\xi)$ for the observer with the proper vector (\ref{xi_static}) and present them here. Considering the solution with (\ref{odd_AdS_V}) and the background with (\ref{odd_AdS}) we have to substitute $\bar f$ from (\ref{odd_AdS}) and $f$ from (\ref{odd_AdS_V}) and their derivatives:
\bea
f'(r,v) &=& \frac{1}{p\alf^{2-2/p}}\l(\frac{1}{\l|\ell\r|^2} - \frac{ m_0 + m(v)+ {\cal C}(v)\Theta(r)}{r^{D-1}}\r)^{1/p-1}\nonumber\\
&\times & \l[\frac{2pr}{\l|\ell\r|^2} + (D-2p-1)\frac{ m_0 + m(v)+ {\cal C}(v)\Theta(r)}{r^{D-2}}- \frac{{\cal C}(v)\Theta'(r)}{r^{D-3}} \r];\m{f_prime}\\
\bar f'(r) &=& \frac{1}{p\alf^{2-2/p}}\l(\frac{1}{\l|\ell\r|^2} - \frac{ m_0 }{r^{D-1}}\r)^{1/p-1}
 \l[\frac{2pr}{\l|\ell\r|^2} + (D-2p-1)\frac{ m_0 }{r^{D-2}}\r];\m{f_bar_prime}\\
\dot f(r,v) &=& -\frac{1}{p\alf^{2-2/p}}\l(\frac{1}{\l|\ell\r|^2} - \frac{ m_0 + m(v)+ {\cal C}(v)\Theta(r)}{r^{D-1}}\r)^{1/p-1}\frac{\dot m(v)+ \dot{\cal C}(v)\Theta(r)}{r^{D-3}}
\m{f_dot}
\eea
into (\ref{sup_tau_1})-(\ref{10_CL_1}). Due to conditions for a freely falling observer in subsection \ref{AdS_freely} one has
\be
\frac{m_0}{r^{D-1}}-\frac{1}{|\ell|^2} \geq 0\,.
\m{000}
\ee
As a result, we obtain for the energy density measured by the observer:
\bea
&&{\cal I}^0(\xi) = (D-2)\frac{\sqrt{q}}{2\k}r^{D-3}\l[{1+\frac{r}{\alf^{1-1/p}}\l(\frac{m_0}{r^{D-1}}-\frac{1}{|\ell|^2} \r)^{1/2p}}\r]^{-1}\nonumber\\
&\times& \l\{\l[\l(1+ \frac{m(v)+ {\cal C}(v)\Theta(r)}{r^{D-1}}\l(\frac{m_0}{r^{D-1}}-\frac{1}{|\ell|^2} \r)^{-1} \r)^{1/p}-1 \r]\l[(D-2p-1) \l(\frac{pm_0}{r^{D-2}}-\frac{pr}{|\ell|^2} \r) \r.\r. \nonumber\\
&+& \l. \l. \l((D-2p-1)\frac{m_0}{r^{D-2}}+\frac{2pr}{|\ell|^2} \r)\l(p - \frac{3}{2} +\frac{1}{2}\l[{1+\frac{r}{\alf^{1-1/p}}\l(\frac{m_0}{r^{D-1}}-\frac{1}{|\ell|^2} \r)^{1/2p}}\r]^{-1}\r)\r]\r. \nonumber\\
&-&  \l.(D-2p-1)\frac{m_0}{r^{D-2}}-\frac{2pr}{|\ell|^2} +
\l(1+ \frac{m(v)+ {\cal C}(v)\Theta(r)}{r^{D-1}}\l(\frac{m_0}{r^{D-1}}-\frac{1}{|\ell|^2} \r)^{-1} \r)^{1/p-1}\r.\nonumber
\eea
\be
\times \l.\l[ (D-2p-1)\frac{m_0 + m(v)+ {\cal C}(v)\Theta(r)}{r^{D-2}} - \frac{{\cal C}(v)\Theta'(r)}{r^{D-3}} +\frac{2pr}{|\ell|^2}\r] \r\}\,.
\m{I_0_AdS}
\ee
For the flux density measured by the freely falling observer one has
\bea
&&{\cal I}^1(\xi) = -(D-2)\frac{\sqrt{q}}{2\k}\l[{1+\frac{r}{\alf^{1-1/p}}\l(\frac{m_0}{r^{D-1}}-\frac{1}{|\ell|^2} \r)^{1/2p}}\r]^{-1}\nonumber\\
&\times& \l(1+ \frac{m(v)+ {\cal C}(v)\Theta(r)}{r^{D-1}}\l(\frac{m_0}{r^{D-1}}-\frac{1}{|\ell|^2} \r)^{-1} \r)^{1/p-1}\l(\dot m(v)+ \dot{\cal C}(v)\Theta(r) \r) \,.
\m{I_1_AdS}
\eea
\subsection{The case of asymptotically dS black holes}
\m{case_dS}

Here, we give the local energetic characteristics represented by the components of the current ${\cal I}^\alf(\xi)$ for the observer with the proper vector (\ref{xi_static}) and with (\ref{const_2}) on the background of static asymptotically dS black hole. Consider the solution with (\ref{dS_5}) and the background with (\ref{dS_1}), substitute $\bar f$ from (\ref{dS_1}) and $f$ from (\ref{dS_5}) and their derivatives:
\bea
f'(r,v) &=& -\frac{1}{p\alf^{2-2/p}}\l(\frac{1}{\ell^2} + \frac{ m_0 + m(v)+ {\cal C}(v)\Theta(r)}{r^{D-1}}\r)^{1/p-1}\nonumber\\
&\times & \l[\frac{2pr}{\ell^2} - (D-2p-1)\frac{ m_0 + m(v)+ {\cal C}(v)\Theta(r)}{r^{D-2}}+ \frac{{\cal C}(v)\Theta'(r)}{r^{D-3}} \r];\m{f_prime_dS}\\
\bar f'(r) &=& -\frac{1}{p\alf^{2-2/p}}\l(\frac{1}{\ell^2} + \frac{ m_0 }{r^{D-1}}\r)^{1/p-1}
 \l[\frac{2pr}{\ell^2} - (D-2p-1)\frac{ m_0 }{r^{D-2}}\r];\m{f_bar_prime_dS}\\
\dot f(r,v) &=& -\frac{1}{p\alf^{2-2/p}}\l(\frac{1}{\ell^2} + \frac{ m_0 + m(v)+ {\cal C}(v)\Theta(r)}{r^{D-1}}\r)^{1/p-1}\frac{\dot m(v)+ \dot{\cal C}(v)\Theta(r)}{r^{D-3}}
\m{f_dot_dS}
\eea
 into (\ref{sup_tau_1})-(\ref{10_CL_1}). As a result we obtain the energy density:
\bea
&&{\cal I}^0(\xi) = (D-2)\frac{\sqrt{q}}{2\k}r^{D-3}|c|\l[{1+\l(1- |c|+ \frac{|c|r^2}{\alf^{2-2/p}}\l(\frac{m_0}{r^{D-1}}+\frac{1}{\ell^2} \r)^{1/p}\r)^{1/2}}\r]^{-1}\nonumber\\
&\times& \l\{\l[\l(1+ \frac{m(v)+ {\cal C}(v)\Theta(r)}{r^{D-1}}\l(\frac{m_0}{r^{D-1}}+\frac{1}{\ell^2} \r)^{-1} \r)^{1/p}-1 \r] \r. \nonumber\\
&\times& \l.\l[ (D-2p-1) \l(\frac{pm_0}{r^{D-2}}+\frac{pr}{\ell^2} \r)+ \l(p - 1\r) \l((D-2p-1)\frac{m_0}{r^{D-2}}-\frac{2pr}{\ell^2} \r)\r]  \r. \nonumber\\
&+& \l. \frac{|c|r^2}{2\alf^{2-2/p}} \l((D-2p-1)\frac{m_0}{r^{D-2}}-\frac{2pr}{\ell^2} \r)\, \l(\frac{m_0}{r^{D-1}}+\frac{1}{\ell^2} \r)^{1/p}\r. \nonumber\\
&\times& \l.\l[{\l(1- |c|+ \frac{|c|r^2}{\alf^{2-2/p}}\l(\frac{m_0}{r^{D-1}}+\frac{1}{\ell^2} \r)^{1/p}\r)+\l(1- |c|+ \frac{|c|r^2}{\alf^{2-2/p}}\l(\frac{m_0}{r^{D-1}}+\frac{1}{\ell^2} \r)^{1/p}\r)^{1/2}}\r]^{-1}\r.\nonumber\\
&\times& \l.\l[\l(1+ \frac{m(v)+ {\cal C}(v)\Theta(r)}{r^{D-1}}\l(\frac{m_0}{r^{D-1}}+\frac{1}{\ell^2} \r)^{-1} \r)^{1/p}-1 \r] \r.\nonumber\\
&-&  \l.(D-2p-1)\frac{m_0}{r^{D-2}}+\frac{2pr}{\ell^2} +
\l(1+ \frac{m(v)+ {\cal C}(v)\Theta(r)}{r^{D-1}}\l(\frac{m_0}{r^{D-1}}+\frac{1}{\ell^2} \r)^{-1} \r)^{1/p-1}\r.\nonumber\\
&\times& \l.\l[ (D-2p-1)\frac{m_0 + m(v)+ {\cal C}(v)\Theta(r)}{r^{D-2}} - \frac{{\cal C}(v)\Theta'(r)}{r^{D-3}} -\frac{2pr}{\ell^2}\r] \r\}\,.
\m{I_0_dS}
\eea
The flux density measured by the freely falling observer is
\bea
&&{\cal I}^1(\xi) = - (D-2)\frac{\sqrt{q}}{2\k}r^{D-3}|c|\l[{1+\l(1- |c|+ \frac{|c|r^2}{\alf^{2-2/p}}\l(\frac{m_0}{r^{D-1}}+\frac{1}{\ell^2} \r)^{1/p}\r)^{1/2}}\r]^{-1}\nonumber\\
&\times& \l(1+ \frac{m(v)+ {\cal C}(v)\Theta(r)}{r^{D-1}}\l(\frac{m_0}{r^{D-1}}+\frac{1}{\ell^2} \r)^{-1} \r)^{1/p-1}\l(\dot m(v)+ \dot{\cal C}(v)\Theta(r) \r) \,.
\m{I_1_dS}
\eea

\subsection{The case of black holes with flat asymptotics}
\m{case_flat}

 Here, we present local energetic characteristics ${\cal I}^\alf(\xi)$ for freely falling observer on the background of a static black hole with flat asymptotics and with the proper vector (\ref{xi_static}). Considering the solution with (\ref{F_3}) and the background with (\ref{F_1}), we substitute $\bar f$ from (\ref{F_1}) and $f$ from (\ref{F_3}) and their derivatives:
\bea
f'(r,v) &=& \frac{1}{p\alf^{2-2/p}}\l(\frac{ m_0 + m(v)+ {\cal C}(v)\Theta(r)}{r^{D-2p-1}}\r)^{1/p-1}\nonumber\\
&\times & \l[(D-2p-1)\frac{ m_0 + m(v)+ {\cal C}(v)\Theta(r)}{r^{D-2p}}- \frac{{\cal C}(v)\Theta'(r)}{r^{D-2p-1}} \r];\m{f_prime_F}
\eea
\bea
\bar f'(r) &=& \frac{1}{p\alf^{2-2/p}}\l(\frac{ m_0 }{r^{D-2p-1}}\r)^{1/p-1}
 \l[(D-2p-1)\frac{ m_0 }{r^{D-2p}}\r];\m{f_bar_prime_F}\\
\dot f(r,v) &=& -\frac{1}{p\alf^{2-2/p}}\l(\frac{ m_0 + m(v)+ {\cal C}(v)\Theta(r)}{r^{D-2p-1}}\r)^{1/p-1}\frac{\dot m(v)+ \dot{\cal C}(v)\Theta(r)}{r^{D-2p-1}}
\m{f_dot_F}
\eea
 into (\ref{sup_tau_1})-(\ref{10_CL_1}). Finally we obtain the energy density measured by the observer:
\bea
&&{\cal I}^0(\xi) = (D-2)\frac{\sqrt{q}}{2\k}r^{D-3}\l[{1+\frac{r}{\alf^{1-1/p}}\l(\frac{m_0}{r^{D-1}} \r)^{1/2p}}\r]^{-1}\nonumber\\
&\times& \l\{\l[\l(1+ \frac{m(v)+ {\cal C}(v)\Theta(r)}{m_0} \r)^{1/p}-1 \r]\r. \nonumber\\
&\times&  \l. (D-2p-1)\frac{m_0}{r^{D-2}}\l(2p - \frac{3}{2} +\frac{1}{2}\l[{1+\frac{r}{\alf^{1-1/p}}\l(\frac{m_0}{r^{D-1}} \r)^{1/2p}}\r]^{-1}\r)\r. \nonumber\\
&-&  \l.(D-2p-1)\frac{m_0}{r^{D-2}} +
\l(1+ \frac{m(v)+ {\cal C}(v)\Theta(r)}{m_0}\r)^{1/p-1}\r.\nonumber\\
&\times& \l.\l[ (D-2p-1)\frac{m_0 + m(v)+ {\cal C}(v)\Theta(r)}{r^{D-2}} - \frac{{\cal C}(v)\Theta'(r)}{r^{D-3}}\r] \r\}\,.
\m{I_0_flat}
\eea
For the flux density measured by the freely falling observer one has
\bea
&&{\cal I}^1(\xi) = -(D-2)\frac{\sqrt{q}}{2\k}\l[{1+\frac{r}{\alf^{1-1/p}}\l(\frac{m_0}{r^{D-1}} \r)^{1/2p}}\r]^{-1}\nonumber\\
&\times& \l(1+ \frac{m(v)+ {\cal C}(v)\Theta(r)}{m_0} \r)^{1/p-1}\l(\dot m(v)+ \dot{\cal C}(v)\Theta(r) \r) \,.
\m{I_1_flat}
\eea

\ed
\begin{thebibliography}{999}


\bibitem{Lovelock} Lovelock D 1971 The Einstein tensor and its
generalizations {\em J. Math. Phys.} {\bf 12} 498

\bibitem{Gross_Witten_1986} Gross D J and Witten E 1986 Superstring modifications of Einstein's equations {\em Nucl. Phys. B} {\bf 277} 1-10

\bibitem{Bento_Bertolami_1996} Bento M C and Bertolami O 1996 Maximally symmetric cosmological solutions of higher-curvature string effective theories with dilatons
   {\em Phys. Lett. B} {\bf 368} 198-201


\bibitem{Troncoso+_2000} Crisostomo J, Troncoso R and Zanelli J 2000 Black Hole Scan {\em Phys. Rev. D} {\bf 62} 084013   	arXiv:hep-th/0003271

\bibitem{Kastor_Mann_2006} Kastor D and Mann R 2006 On black strings and branes in Lovelock gravity {\em JHEP0406} {\bf 2006} 048  	arXiv:hep-th/0603168

\bibitem{Giribet_Oliva_Troncoso_2006} Giribet G, Oliva J and Troncoso R 2006 Simple compactifications and Black p-branes in Gauss-Bonnet and Lovelock Theories {\em JHEP0605} {\bf 2006} 007 arXiv:hep-th/0603177

\bibitem{Cai_Ohta_2006} Cai R-G and Ohta N 2006 Black Holes in Pure Lovelock Gravities {\em Phys. Rev. D} {\bf 74}  064001  arXiv:hep-th/0604088

\bibitem{Cai+_2008} Cai R-G, Cao L-M, Hu Y-P and Kim S P 2008 Generalized Vaidya Spacetime in Lovelock Gravity and Thermodynamics on Apparent Horizon {\em Phys. Rev. D} {\bf 78}  124012  arXiv:0810.2610 [hep-th]

\bibitem{Dadhich_2010} Dadhich N 2010 Characterization of the Lovelock gravity by Bianchi derivative {\em Pramana} {\bf 74}  875-882 arXiv:0802.3034 [gr-qc]

\bibitem{Dadhich+_2012a} Dadhich N, Ghosh S G  and Jhingan S 2012 The Lovelock gravity in the critical spacetime dimension {\em Phys. Lett. B} {\bf 711} 196 arXiv:1202.4575 [gr-qc]

\bibitem{Dadhich+_2013a} Dadhich N, Ghosh S G  and Jhingan S 2013 Gravitational collapse in pure Lovelock gravity in higher dimensions
 {\em Phys Rev. D} {\bf 88} 084024  arXiv:1308.4312 [gr-qc]

\bibitem{Dadhich+_2013b} Dadhich N, Ghosh S G  and Jhingan S 2013 Bound orbits and gravitational theory {\em Phys. Rev. D} {\bf 88} 124040 arXiv:1308.4770 [gr-qc]

\bibitem{Dadhich_2016} Dadhich N 2016 A discerning gravitational property for gravitational equation in higher dimensions
{\em Euro. Phys. J. C} {\bf 76} 104 arXiv:1506.08764 [gr-qc]


\bibitem{Dadhich_2011} Dadhich N 2011 On Lovelock vacuum solution {\em Math.Today} {\bf 26} 37 arXiv:1006.0337 [hep-th]

\bibitem{Dadhich+_2012} Dadhich N, Pons J M and Prabhu 2012 Thermodynamical universality of the Lovelock black holes  {\em Gen. Relat. Grav.} {\bf 44} 2595-2601 arXiv:1110.0673 [gr-qc]

\bibitem{Dadhich+_2013} Dadhich N, Pons J M and Prabhu 2013 On the static Lovelock black holes   {\em Gen. Relat. Grav.} {\bf 45}  1131-1144 arXiv:1201.4994 [gr-qc]


\bibitem{F_2020} Forghani S D, Mazharimousavi D S and Halilsoy M 2020 Higher Dimensional Particle Model in Pure Lovelock Gravity
arXiv:2002.12753 [physics.gen-ph]

\bibitem{Dadhich+_2020} Chakraborty S and Dadhich N 2020 Limits on stellar structures in Lovelock theories of gravity {\em Phys. Dark Univ.} {\bf 30} 100658 arXiv:2005.07504 [gr-qc]

\bibitem{Dadhich+_2020a} Shaymatov S and Dadhich N 2020 Weak cosmic censorship conjecture in the pure Lovelock gravity arXiv:2008.04092 [gr-qc]

\bibitem{Kastikainen_2020} Kastikainen J 2020 Quasi-local energy and ADM mass in pure Lovelock gravity {\em Class. Quantum Grav.} {\bf 37} 025001 arXiv:1908.05522 [gr-qc]

\bibitem{Deser_Franklin} Deser S and Franklin J 2012 Canonical Analysis and Stability of Lanczos-Lovelock Gravity {\em Class. Quantum Grav.} {\bf 29} 072001 arXiv:1110.6085 [gr-qc]

\bibitem{Petrov_2019} Petrov A 2019 Field-theoretical construction of currents and superpotentials in Lovelock gravity {\em Class. Quantum Grav.} {\bf 36} 235021 arXiv:1903.05500 [gr-qc]


\bibitem{GPP} Grishchuk L P, Petrov A N  and Popova A D 1984
Exact theory of the (Ein\-stein) gravitational field in an arbitrary
background space-time {\em Commun. Math. Phys.} {\bf 94} 379

\bibitem{[15]} Popova A D and   Petrov A N 1988
The  dynamic  theories  on  a fixed  background in gravitation {\em
Int. J. Mod. Phys. A} {\bf 3} 2651

\bibitem{GP87} Grishchuk L P and Petrov A N 1987 The Hamiltonian description
of the gravitational field and gauge symmertries {\em Sov. Phys.:
JETP} {\bf 65} 5 [1986 {\em Zh. Eksp. Teor. Fiz.} {\bf 92}, 9]

\bibitem{CQG_DT}  Petrov A N 2005 A note on the Deser-Tekin charges {\em Class. Quantum Grav.} {\bf 22} L83 arXiv:gr-qc/0504058

\bibitem{Petrov_2008} Petrov A N 2008 Nonlinear perturbations and conservation laws on
curved backgrounds in GR and other metric theories, 2-nd chapter in
the book: {\em Classical and Quantum Gravity Research} Eds:
Christiansen M N and Rasmussen T K (Nova Science Publishers, N.Y.)
79 arXiv:0705.0019 [gr-qc]

\bibitem{Petrov+_2017}	Petrov A N, Kopeikin S M, Lompay R R and Tekin B 2017
{\em Metric Theories of Gravity: Perturbations and Conservation Laws} (de Gruyter: Germany)

\bibitem{Petrov_Pitts_2019}  Petrov A N and Pitts J B 2019 The Field-Theoretic Approach in General Relativity and Other Metric Theories. A Review  {\em Space, Time and Fundamental Interactions} {\bf no. 4} 66-124 arXiv:2004.10525 [gr-qc]



\bibitem{Krishna_2019} Bhattacharya K and Majhi B R 2019 Abbott-Deser-Tekin like conserved quantities in Lanczos-Lovelock gravity: beyond Killing diffeomorphisms {\em Class. Quantum Grav.} {\bf 36} 065009 arXiv:1806.05519 [gr-qc]

\bibitem{DT1} Deser S and Tekin B 2002 Gravitational energy in
quadratic curvature gravities {\it Phys. Rev. Lett.} {\bf 89} 101101 arXiv:hep-th/0205318

\bibitem{DT2} Deser S and Tekin B 2003 Energy in generic
higher curvature gravity theories {\em Phys. Rev. D} {\bf 67} 084009 arXiv:hep-th/0212292

\bibitem{DT_2019} Adami H, Setare M R, Tahsin T C and Tekin B 2019 Conserved Charges in Extended Theories of Gravity {\em  Physics Reports}  {\bf 834} 1 arXiv:1710.07252 [hep-th]


\bibitem{Nozawa_Maeda_2006} Nozawa M and Maeda H 2006 Effects of Lovelock terms on the final fate of gravitational collapse: analysis in dimensionally continued gravity  {\em Class. Quantum Grav.} {\bf 23} 1779 arXiv:gr-qc/0510070

\bibitem{Dominguez_Gallo_2006} Dominguez A E and Gallo E 2006 Radiating black hole solutions in Einstein-Gauss-Bonnet gravity {\em Phys. Rev. D} {\bf 73} 064018 arXiv:gr-qc/0512150

\bibitem{Hawking_Ellis_1973} Hawking S W and Ellis G F R 1973 {\it The large scale structure of space-time} (CUP: Cambridge)

\bibitem{Ashtekar_Krishnan_2004} Ashtekar A and Krishnan B 2004 Isolated and dynamical horizons and their applications {\em	Living Rev. Rel.} {\bf 7} 10  	arXiv:gr-qc/0407042

\bibitem{Booth_2005} Booth I 2005 Black hole boundaries {\em Can. J. Phys.} {\bf 83} 1073-1099 arXiv:gr-qc/0508107

\bibitem{Nielsen_2010} Nielsen A B 2010 The spatial relation between the event horizon and trapping horizon  {\em Class. Quantum Grav.} {\bf 27} 245016  arXiv:1006.2448 [gr-qc]

\bibitem{Nielsen+_2011} Nielsen A B, Jasiulek M, Krishnan B and Schnetter E 2011 The slicing dependence of non-spherically symmetric quasi-local horizons in Vaidya spacetimes  	{\em Phys. Rev. D} {\bf 83} 124022 arXiv:1007.2990 [gr-qc]

\bibitem{AFrancavigliaR} Allemandi G, Francaviglia M and Raiteri M
2003 Charges and energy in Chern-Simons theories and Lovelock
gravity {\em Class. Quantum Grav}. {\bf 20} 5103 arXiv:gr-qc/0308019

\bibitem{Okuyama} Okuyama N and Koga J-I 2005 Asymptotically anti-de
Sitter spacetimes and conserved quantities in higher curvature
gravitational theories {\em Phys. Rev. D} {\bf 71}  084009  arXiv:hep-th/0501044

\bibitem{DerKatzOgushi} Deruelle N, Katz J and Ogushi S 2004 Conserved
charges in Einstein-Gauss-Bonnet theory {\it Class. Quantum Grav.}
{\bf 21} 1971 arXiv:gr-qc/0310098

\bibitem{Kofinas_Olea_2007} Kofinas G and Olea R 2007 Universal regularization prescription for Lovelock AdS gravity
{\em  JHEP} {\bf 0711} 069 arXiv:0708.0782 [hep-th]

\bibitem{Peng_Liu_2020} Peng J-J and Liu H-F 2020 A new formula for conserved charges of Lovelock gravity in AdS space–times and its generalization {\em Int. J. Mod. Phys. A} {\bf 35} 2050102  	arXiv:1912.08013 [gr-qc]

\bibitem{Misner_Thorn_Wheeler_1973} Misner C W,  Thorne K S and Wheeler J A 1973 {\it Gravitation} (W H Freeman and Company: San Francisco)

\bibitem{Brown_York_1993} Brown J D and York J W 1993 Quasilocal energy and conserved charges derived from the gravitational action {\em Phys. Rev. D} {\bf 47} 1407-1419 arXiv:gr-qc/9209012


\bibitem{Nariai_1950}  Nariai H 1950 On some static solutions of Einstein’s gravitational field equations in a spherically symmetric case {\em Sci. Rep. Tohoku Univ.} {\bf 34} 160

\bibitem{Nariai_1951}  Nariai H 1951 On a new cosmological solution of Einstein’s field equations of gravitation  {\em Sci. Rep. Tohoku Univ.} {\bf 35} 62

\bibitem{dS_1}   Balasubramanian V, de Boer J and Minic D 2002 Mass, Entropy and Holography in Asymptotically de Sitter Spaces {\em Phys. Rev. D} {\bf 65} 123508 arXiv:hep-th/0110108
\bibitem{dS_2}  Cai R-G, Myung Y S and Zhang Y-Z 2002 Check of the Mass Bound Conjecture in the de Sitter Space {\em Phys. Rev. D} {\bf 65} 084019 arXiv:hep-th/0110234


\bibitem{Rodrigo_2017}   Arenas-Henriquez G, Miskovic O and Olea O 2017 Vacuum Degeneracy and Conformal Mass in Lovelock AdS Gravity {\em JHEP} {\bf 1711} arXiv:1710.08512 [hep-th]

\bibitem{Rodrigo+_2020}   Giribet G, Miskovic O, Olea R and Rivera-Betancour D 2020 Topological invariants and the definition of energy in quadratic gravity theory {\em Phys. Rev. D} {\bf 101} 064046 arXiv:2001.09459 [hep-th]

\bibitem{Olea_2005} Olea R 2005   Mass, Angular Momentum and Thermodynamics in Four-Dimensional Kerr-AdS Black Holes
{\em JHEP} {\bf 0506} 023 arXiv:hep-th/0504233

\bibitem{Olea_2007} Olea R 2007 Regularization of odd-dimensional AdS gravity: Kounterterms
{\em JHEP} {\bf 0704} 073 arXiv:hep-th/0610230

\bibitem{Miskovic_Olea_2007} Miskovic O and Olea R 2007 Counterterms in Dimensionally Continued AdS Gravity
{\em JHEP} {\bf 0710} 028 arXiv:0706.4460 [hep-th]


\bibitem{Szabados_2009} Szabados L 2009 Quasi-Local Energy-Momentum and Angular Momentum in General Relativity {\em Living Reviews in Relativity} {\bf 12}, p.p. {163}; {http://www.livingreviews.org/lrr-2009-4}

\bibitem{Chakraborty_Dadhich_2012} Chakraborty S and Dadhich N 2015  Brown-York quasilocal energy in Lanczos-Lovelock gravity and black hole horizons  	{\em JHEP} {\bf 2015} 1-19  	arXiv:1509.02156 [gr-qc]

\bibitem{Misner_Sharp_1964} Misner C W and Sharp D H 1964 Relativistic Equations for Adiabatic, Spherically Symmetric Gravitational Collapse {\em Phys. Rev. B} {\bf 136} 571

\bibitem{Maeda_2006} Maeda H 2006 Final fate of spherically symmetric gravitational collapse of a dust cloud in Einstein-Gauss-Bonnet gravity {\em Phys. Rev. D} {\bf 73} 104004
arXiv:gr-qc/0602109

\bibitem{Maeda_Nozawa_2008} Maeda H and Nozawa M 2008 Generalized Misner-Sharp quasi-local mass in Einstein-Gauss-Bonnet gravity  {\em Phys. Rev. D} {\bf 77} 064031  	arXiv:0709.1199 [hep-th]

\bibitem{Nozawa_Maeda_2008} Nozawa M and Maeda H 2008 Dynamical black holes with symmetry in Einstein-Gauss-Bonnet gravity {\em Class. Quantum Grav.} {\bf 25} 055009 arXiv:0710.2709 [gr-qc]

\bibitem{Maeda+_2011} Maeda H, Willison S and Ray S 2011  Lovelock black holes with maximally symmetric horizons  {\em Class. Quantum Grav.} {\bf 28} 165005 arXiv:1103.4184 [gr-qc]










\end{thebibliography}
